\documentclass{article}
\usepackage{epstopdf}
\usepackage[english]{babel}
\usepackage[T1]{fontenc}
\usepackage[utf8]{inputenc}
\usepackage{eurosym}
\usepackage{textcomp}
\usepackage{calc}
\usepackage{amssymb}
\usepackage{amsmath}
\usepackage{graphicx}
\usepackage{amsthm}

\usepackage{multicol} 
\usepackage{color}

\usepackage{tikz}

\usepackage{calrsfs}
\DeclareMathAlphabet{\pazocal}{OMS}{zplm}{m}{n}

\newcommand{\DDD}{\pazocal{D}}
\newcommand{\EEE}{\pazocal{E}}

\newcommand{\III}{\pazocal{I}}

\newcommand{\PPP}{\pazocal{P}}

\newcommand{\TTT}{\pazocal{T}}

\newsavebox\CBox

\newcommand\hcancel[2][0.5pt]{%
  \ifmmode\sbox\CBox{$#2$}\else\sbox\CBox{#2}\fi%
  \makebox[0pt][l]{\usebox\CBox}%
  \rule[0.5\ht\CBox-#1/2]{\wd\CBox}{#1}}
  
\newcommand{\R}{\mathbb{R}}

\newcommand{\Z}{\mathbb{Z}}

\renewcommand{\P}{\mathbb{P}}
\usepackage{bbm}

\renewcommand{\a}{\alpha}
\renewcommand{\b}{\beta}
\newcommand{\g}{\gamma}

\renewcommand{\d}{\delta}

\newcommand{\e}{\varepsilon}

\renewcommand{\th}{\theta}

\newcommand{\s}{\sigma}

\renewcommand{\phi}{\varphi}

\newcommand{\noi}{\noindent}

\usepackage[toc,page]{appendix}

\usepackage{comment}

\usepackage{eurosym}

\usepackage{relsize} 

\usepackage{mathtools}

\usepackage{longtable}

\usepackage[subnum]{cases}

\usepackage{natbib}

\usepackage{float}

\makeatletter
\newcommand{\vast}{\bBigg@{3}}
\newcommand{\Vast}{\bBigg@{5}}
\makeatother

\usepackage{fullpage}
\usepackage{authblk}

\title{Temporal evolution of the extreme excursions of multivariate $k$th order Markov processes with application to oceanographic data}
\author[1]{Stan Tendijck}
\author[1,2]{Philip Jonathan}
\author[3]{David Randell}
\author[1]{Jonathan Tawn}
\affil[1]{Department of Mathematics and Statistics, Lancaster University LA1 4YW, United Kingdom}
\affil[2]{Shell Research Limited, London SE1 7NA, United Kingdom}
\affil[3]{Shell Global Solutions International B.V., 1031 HW Amsterdam, Netherlands}

\begin{document}
\maketitle




%

\section*{Abstract}
We develop two models for the temporal evolution of extreme events of multivariate $k$th order Markov processes. The foundation of our methodology lies in the conditional extremes model of \citet{heffernan2004conditional}, and it naturally extends the work of \citet{winter2016,winter2017} and \citet{tendijck2019} to include multivariate random variables. We use cross-validation-type techniques to develop a model order selection procedure, and we test our models on two-dimensional meteorological-oceanographic data with directional covariates for a location in the northern North Sea. We conclude that the newly-developed models perform better than the widely used historical matching methodology for these data.

\bigskip

\noi\textbf{Keywords:} {\em extreme value theory, time-series, Markov processes, oceanography}

\section{Introduction}\label{paper4::sec::myintro}
Farmers, stock brokers and sailors have one thing in common: they or their businesses are most heavily affected by extreme events like droughts and rainfall, stock market crashes, or extreme winds and waves, respectively. Understanding the statistical behaviour of such events as a whole is crucial for risk analyses. 
To make this more precise, if we let $(\bold{X}_t)_{t\in\Z}$ be a stationary $d$-dimensional random process of interest, then we seek to model excursions of the process in and out of a set $E\subset\R^d$ in time, i.e., the behaviour of
\begin{equation}\label{paper4::def::cluster}
\{\bold{X}_i:\ i = a,\dots,b;\ \bold{X}_{i} \in E;\ \bold{X}_{a-1},\bold{X}_{b+1} \not\in E\},
\end{equation}
where $E$ is associated with extreme events of the random variable $\bold{X}$ which is identically distributed to any $\bold{X}_j$, $j\in\Z$. Moreover, we assume that the random process consists of multiple components that can be extreme. To solve this task, we assume that the multivariate random process is a realisation of a $k$th order Markov chain.

We use extreme value theory, a subfield of statistics, to characterise excursions. There is considerable attention to this area in the literature, 
but most of extreme value theory for stationary Markov chains dates back over $20$ years. \citet{rootzen1988} and \citet{perfekt1997} develop limiting results for univariate Markov chains and multivariate Markov chains, respectively. \citet{smith1992} calculates the extremal index \citep{leadbetter1983} for a univariate Markov chain and \citet{smith1997} use parametric bivariate transition distributions to model the extremes of a univariate first order Markov process. Finally, \citet{yun2000} develops asymptotic theory for functionals of univariate $k$th order Markov extreme events. All of these authors derive results under the assumption of asymptotic dependence \citep{joe1997}, i.e., for a stationary process $(X_t)_{t\in\Z}$ satisfying suitable long-range mixing conditions, under the assumption that for any lag $l=1,2,\dots$
\[
\lim_{u\to x^*} \P(X_{t+l}>u|X_t>u) > 0,
\]
where $x^*$ is the right upper end point of the distribution of $X_t$. This early work doesn't consider what happens when asymptotic independence is present, i.e., when this limiting probability converges to $0$ for some $l$. The first paper which considers such processes is \citet{bortot1998} who assume a first order Markov model, with \citet{ledford2003diagnostics} considering a general framework for the modelling of asymptotic independent processes, and key recent probabilistic developments given by \citet{papa2017} and \citet{papa2023}.

\citet{randell2015} speculate that a statistical model for the evolution of (multivariate) trajectories would be a valuable enhancement of description of ocean storm events. The first statistical work the current authors are aware of, that defines a model for the distribution of all observations during an excursion is \citet{winter2016}, who assume a flexible univariate first order Markov process exhibiting either asymptotic independence or asymptotic dependence across lags. \citet{winter2017} incorporate higher order dependence model to give $k$th order Markov processes with $k>1$. Finally, \citet{tendijck2019} extend that model to a $k$th order univariate Markov process with a directional covariate. We remark that their work cannot be considered to model the extremes of bivariate Markov processes since the associated directional covariate does not take on extreme values. \citet{feld2015} use a sophisticated covariate model for the most extreme observation (the most extreme value of the dominant variable) in an excursion, combined with a historical matching approach for the intra-excursion trajectory; in Section~\ref{paper4::sec::model1} we adopt a version of this methodology as a benchmark for our case study. Finally, we mention well-established literature on multivariate time series, e.g., \citet{tiao1989}, which is not directly applicable to modelling environmental extremes because such models are only designed to model typical behaviours. Financial time-series models, e.g., \citet{bauwens2006}, are also not applicable because these are specifically tailored to model data exhibiting volatility, with tail switching during extreme events \citep{bortot2003}.


In this work, we present a natural extension to \citet{tendijck2019} by defining two multivariate $k$th order Markov models that exhibit both asymptotic (in)dependence across variables and/or at some lags. The work is motivated by our case study in which we model excursions of meteorological-oceanographic (met-ocean) data: significant wave height, wind speed, and their associated directions, for a location in the northern North Sea.



We use the following set up. Assume that at each time $t\in\Z$, the distribution of the $d$-dimensional random variable $\bold{X}_t$ is stationary through time; that is, $\bold{X}_t$ has the same distribution as some $\bold{X}=(X_1,\dots,X_d)$ with distribution function  $F_{\bold{X}}$. 
For $1\leq j\leq d$, write $F_{X_j}$ as the $j$th marginal distribution of $F_{\bold{X}}$. The distribution functions $F_{X_j}$ are unknown and must be estimated. For extreme arguments of $F_{X_j}$, we use univariate extreme value theory to motivate a class of parametric tail forms. More precisely, we assume that for each $1\leq j\leq d$, the excesses tail above some high level $u_j\in\R$ of the marginal distribution $F_{X_j}$ are approximated with a generalised Pareto distribution \citep{davison1990}. For non-extreme arguments $x<u_j$ of the function $F_{X_j}$, an empirical model usually suffices. 

In multivariate extreme value theory, it is common to consider the marginals and the dependence of random variables separately, such that the usually-dominant marginal effect does not influence the modelling of a possibly complex dependence structure. So given the marginal models as discussed above, we transform the random process $(\bold{X}_t)_{t\in\Z}$ onto standard Laplace margins $(\bold{Y}_t)_{t\in\Z}$ using the transformation: $X_j\mapsto Y_j:=F_L^{-1}(F_{X_j}(X_j))$, where $F_L^{-1}$ is the inverse of the standard Laplace distribution function. Here the choice of Laplace margins is made to allow for the modelling of potential negative dependence at certain lags or across components \citep{keef2013estimation}.

For multivariate random processes, there are many ways of defining an extreme event. In our case study, we take the met-ocean variable significant wave height $H_S$ as the excursion-defining component. We follow \citet{winter2017} and \citet{tendijck2019} in adopting the conditional extremes model of \citet{heffernan2004conditional}, see also Section~\ref{paper4::sec::model0}, as the foundation of our approach. Without loss of generality, we first define the component $X_1$ of $\bold{X}$ as the defining variable for the extreme events. So, we set our excursion set 
$E=E_u:=(F_{X_1}^{-1}\{F_L(u)\},\infty)\times \R^{d-1}$ for some high threshold $u\in\R_+$ 
and rewrite our definition of an excursion as
\begin{equation}\label{paper4::def::cluster}
\{\bold{Y}_i:\ i=a,\dots,b;\ Y_{i,1} > u; Y_{a-1,1} \leq u,\ Y_{b+1,1}\leq u\}
\end{equation}
for $a,b\in\Z$, indices for the start and the end time points of the excursion, respectively. In shorthand, the excursion is then $\bold{Y}_{a:b}$. We remark that in this definition, we accept that multiple excursions can occur close together in time, and thus these cannot be considered independent. The reason for this choice is that imposing a minimal separation of excursions would complicate the modelling significantly. We recognize that this is a feature of the current approach which can be improved upon in future work.



The remaining part of this paper is organised as follows. 
In Section~\ref{paper4::sec::models}, we present our strategy for modelling excursions by defining time intervals corresponding to so-called ``pre-peak'', ``peak'' and ``post-peak'' periods, and we present our $k$th order Markov models for each of these time periods. 
In Section~\ref{paper4::sec::casestudy}, we apply the two Markov model forms we propose to met-ocean data for a location in the northern North Sea. We compare the model performance with a baseline historical matching approach  by assessing their respective performance in estimating the tails of the distributions of complex structure variables \citep{coles1994statistical}, corresponding to approximations of the response of hypothetical offshore or coastal facilities to extreme met-ocean environments. We find that in general the new models are preferred.


\begin{figure}[!htbp]
\centering\includegraphics[width=0.75\textwidth]{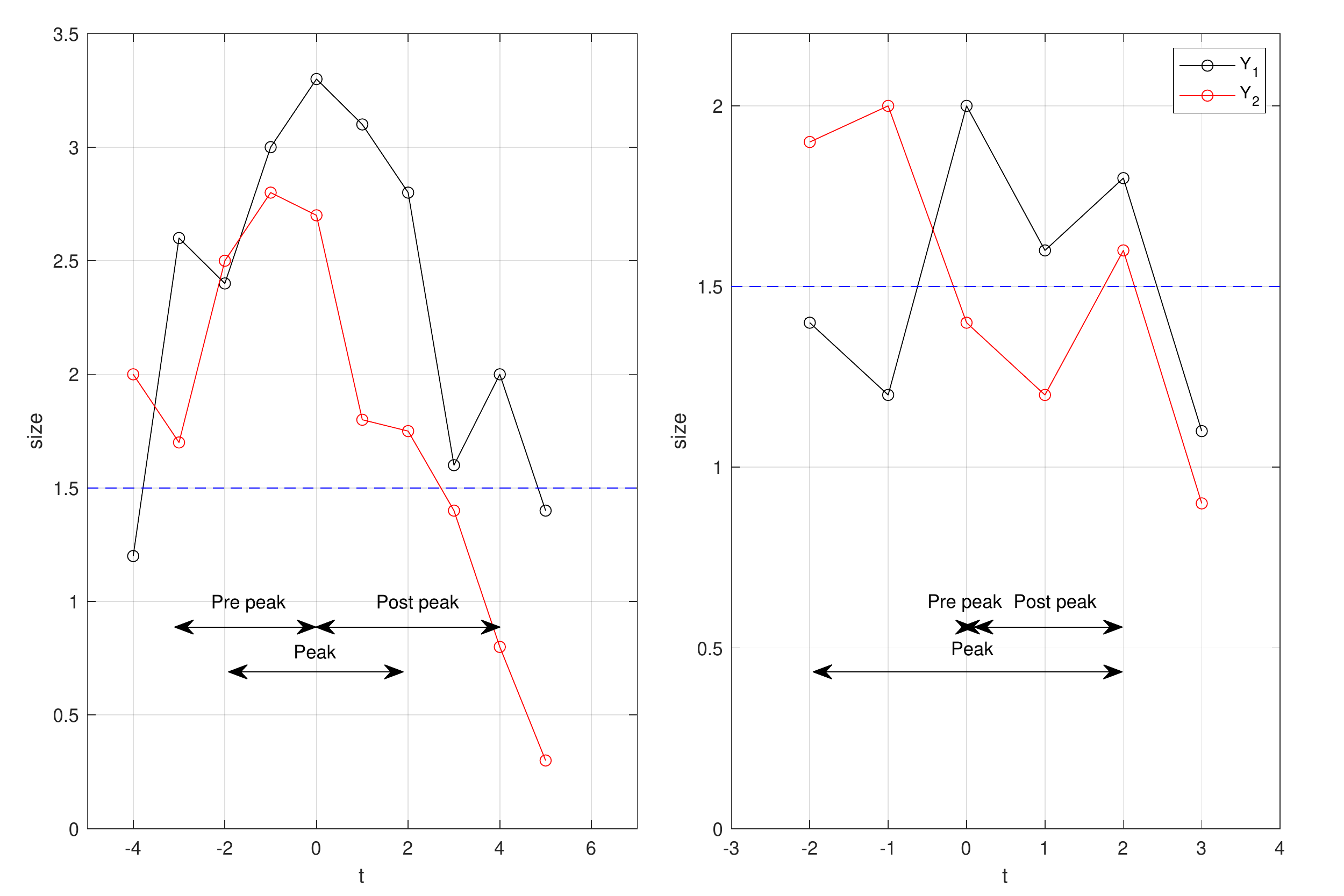}
\caption{Illustration of the periods of the pre-peak, peak and post-peak periods for two excursions from a Markov model with order $k=3$.}
\label{paper4::fig::simillustration}
\end{figure}

\section{The models}\label{paper4::sec::models}

\subsection{Modelling strategy}\label{paper4::sec::modellingstrat}

To model excursions as in definition~(\ref{paper4::def::cluster}), two types of approaches have been proposed in the literature of univariate extremes: a forward model \citep{rootzen1988} and a peak model \citep{smith1997}. Both of these are two-step approaches by nature. The forward model first describes the distribution of a random exceedance $Y_t>u$ with a univariate extremes model and a conditional model for the distribution for any $j\geq 1$ of $Y_{t+j}|(Y_{t+j-i}=y_{t+j-i}\ i=1,\dots,j)$ where $y_{t}>u$. Even though this approach does not directly model the univariate equivalent of excursions in formulation~(\ref{paper4::def::cluster}), 
estimates of some extremal properties of the process $(Y_t)_{t\geq 1}$, such as the extremal index~\citep{leadbetter1983}, can still be obtained by allowing the excursion threshold to be significantly lower than the cluster threshold used in extremal index estimators. Notably, \citet{winter2016,winter2017} use the forward approach in their work.

The peak model, on the other hand, does model excursions as defined here. This method relies on a univariate extremes model for the largest observation of an excursion, e.g., \citet{eastoe2012}, and a conditional model for observations before and after the excursion maximum. 
\citet{winter2016} use this approach for their first order model but not for their $k$th order model \citep{winter2017}. They avoid this method explicitly because of difficulties that arise in preserving model characteristics in forward and backward simulations near the excursion maximum (i.e., the time point at which the defining variate $X_1$ achieves its maximum value during the excursion). 

\citet{tendijck2019} use the peak method, but they do not address the issues associated with forward and backward simulation under the method. Because the excursion maximum is usually the most important observation of an excursion for risk assessments, we also use the peak method in the current work, but with consideration of backward and forward models. We separate the modelling of excursions into three stages: the modelling of the period of the peak, and the modelling of the pre-peak and post-peak periods; see Figure~\ref{paper4::fig::simillustration} in which the three time periods are illustrated for $k=3$. Without loss of generality, let $t=0$ be the time point at which the first component $Y_{t,1}$ takes its maximum value within an excursion such that $Y_{0,1}>u$ for the threshold $u$. The period of the peak  $\PPP^k_0$ of an excursion of a $k$th order model is then defined as the set of $2k-1$ observations: $\PPP^k_0:=\{\bold{Y}_t:\ -(k-1)\leq t\leq k-1\}$ with $Y_{0,1}>u$. The pre-peak $\PPP^{\mathrm{pre}}$ and post-peak $\PPP^{\mathrm{post}}$ periods are defined as the sets of observations that include the excursion maximum and the observations before and after, respectively: 
\[
\PPP^{\mathrm{pre}}:=\{\bold{Y}_t:\ t'\leq t \leq 0,\ \text{with}\ t'=\mathrm{min}\{s<0:\ \min_{i=s,\dots,0}\{Y_{i,1}\}>u\}\}
\]
and 
\[
\PPP^{\mathrm{post}}:=\{\bold{Y}_t:\ 0\leq t\leq t',\ \text{with}\ t'=\mathrm{max}\{s>0:\ \min_{i=0,\dots,s}\{Y_{i,1}\}>u\}\},
\]
so each of them intersects with $\PPP^k_0$.
The length of $\PPP^k_0$ can be longer or shorter than the length of an excursion if the excursion ends within the period of the peak. We choose to define the period $\PPP^k_0$ in this manner so that the pre-peak and post-peak parts of the excursion are both initialized with $k$ observations. 

We then model an excursion as follows: (i) we model the excursion maximum $Y_{0,1}$ using a generalised Pareto distribution; (ii) we model the period of the peak $\PPP^k_0$ conditional on the storm maximum $Y_{0,1}$ using the model described in Section~\ref{paper4::sec::model0}; (iii-a) if $\min_{j=1,\dots,k-1} Y_{j,1} < u$ ($\min_{j=1,\dots,k-1} Y_{-j,1}<u$), 
then the period $\PPP^{\mathrm{post}}$ ($\PPP^{\mathrm{pre}}$) of the excursion has ended; (iii-b) if $\min_{j=1,\dots,k-1} Y_{j,1} \geq u$ ($\min_{j=1,\dots,k-1} Y_{-j,1}\geq u$), 
then the remaining part of the excursion is modelled with our time-series models from Sections~\ref{paper4::sec::model2}-\ref{paper4::sec::model3} until there exist a $j_1,j_2>0$ such that $Y_{j_1,1}<u$ and $Y_{-j_2,1}<u$; (iv) if $\max_{-j_2\leq i\leq j_1} Y_{i,1} > Y_{0,1}$, then the model for the excursion contradicts the definition of the period of the peak of an excursion, and so we reject such occurrences.


In the next sections, we discuss forward models that are applicable to model the post-peak period $\PPP^{\mathrm{post}}$. We model the pre-peak period $\PPP^{\mathrm{pre}}$ using the forward models applied to $(Y_{-t})_{t\in\Z}$ (with potentially different parameters, although these would be the same if the process was time reversible). Importantly, we do not impose consistency in the forward and backward models to yield a $k$th order Markov chain, e.g., in the case of asymptotic dependent Markov chains the precise dependence conditions between the forward and backward hidden tail chains are given by \citet{janssen2014}. We make this choice for two reasons: (i) for environmental applications, such as in this work, the pre-peak and post-peak period have different distributions, see for example the asymmetry in Figure~\ref{paper4::fig::trajectories}, which is due to different physics in the growth and decay of a storm; (ii) the assumption of a $k$th order Markov process is an approximation for the process that generates our data. Thus, imposing forward and backward consistency for a $k$th order Markov chain is likely to yield worse results for our application. So, we consider the violating of this assumption as a benefit more than a limitation as it can yield more flexible descriptions of excursions.

\subsection{The conditional extremes model}\label{paper4::sec::model0}
We introduce the conditional extreme value model of \citet{heffernan2004conditional}, henceforth denoted the HT model, with notation specific to modelling the period of the peak $\PPP^k_0$. The HT model is widely studied and applied to extrapolate tails of multivariate distributions, e.g., in oceanography \citep{ross2020}, finance \citep{hilal2011}, spatio-temporal extremes \citep{simpson2021}, and multivariate spatial extremes \citep{shooter2021_2}. The HT model is a limit model and its form was originally motivated by deriving possible limiting forms for numerous theoretical examples.

Let 
\[
\bold{Y}_{-(k-1):(k-1)}:=\begin{pmatrix} Y_{-(k-1),1} & \cdots & Y_{-(k-1),d} \\ \vdots & & \vdots \\ Y_{k-1,1} & \cdots & Y_{k-1,d} \end{pmatrix}
\]
 be a random matrix on $\R^{(2k-1)\times d}$ with standard Laplace margins \citep{keef2013estimation}, and define the irregular random matrix $\underline{\bold{Y}}$ to be $\bold{Y}_{-(k-1):(k-1)}$ without the $(k,1)$th element $Y_{0,1}$. That is, we define the irregular matrix $\underline{\bold{x}}\in\R^{(2k-1)d-1}$ as follows:
\[
\underline{\bold{x}} = \begin{pmatrix}x_{-k+1,1} & x_{-k+1,2} & \cdots & x_{-k+1,d} \\ \vdots & \vdots & & \vdots \\ x_{-1,1} & x_{-1,2} & \cdots & x_{-1,d} \\  & x_{0,2} & \cdots & x_{0,d} \\ x_{1,1} & x_{1,2} & \cdots & x_{1,d} \\ \vdots & \vdots & & \vdots \\ x_{k-1,1} & x_{k-1,2} & \cdots & x_{k-1,d} \end{pmatrix},
\]
such that $\underline{\bold{x}}$ does not contain the $(k,1)$th element. Equivalently, we can write $\underline{\bold{x}}=\bold{x}_{-(k,1)}$ for $\bold{x}\in \R^{(2k-1) \times d}$. Additionally, we assume that the joint density of $\bold{Y}_{-(k-1):(k-1)}$ exists. 

The conditional extremes model for $\underline{\bold{Y}}$, conditional on $Y_{0,1}$, assumes that irregular parameter matrices $\underline{\boldsymbol{\a}}\in[-1,1]^{(2k-1)d-1}$, $\underline{\boldsymbol{\b}}\in(-\infty,1)^{(2k-1)d-1}$
and a distribution function $H$ with non-degenerate marginals on  $\R^{(2k-1)d-1}$ (the space of irregular matrices) 
exist, such that for all irregular matrices $\underline{\bold{z}}\in \R^{(2k-1)d-1}$ the limit
\[
\lim_{u\to\infty} \P\left(\frac{\underline{\bold{Y}} -\underline{\boldsymbol{\a}} Y_{0,1}}{Y_{0,1}^{\underline{\boldsymbol{\b}}}} \leq \underline{\boldsymbol{z}},\ Y_{0,1}-u>y\ \Bigg|\ Y_{0,1}>u\right)
\]
exists, assuming component-wise operations, and that
\begin{equation}\label{paper4::htlimiteq1}
H(\underline{\bold{z}}) := \lim_{y\to\infty} \P\left(\frac{\underline{\bold{Y}} -\underline{\boldsymbol{\a}} Y_{0,1}}{Y_{0,1}^{\underline{\boldsymbol{\b}}}} \leq \underline{\boldsymbol{z}}\ \bigg|\ Y_{0,1}=y\right)
\end{equation}
exists, where $\a_{i,j}$, $\b_{i,j}$ and $z_{i,j}$ are the $(i,j)$th elements of $\underline{\boldsymbol{\alpha}}$, $\underline{\boldsymbol{\beta}}$ and $\underline{\bold{z}}$, respectively. This then implies, according to l'Hopital's rule, that for $y>0$, $\underline{\boldsymbol{z}}\in \R^{(2k-1)d-1}$
\begin{equation}\label{paper4::htlimiteq2}
\lim_{u\to\infty} \P\left(\frac{\underline{\bold{Y}} -\underline{\boldsymbol{\a}} Y_{0,1}}{Y_{0,1}^{\underline{\boldsymbol{\b}}}} \leq \underline{\boldsymbol{z}},\ Y_{0,1}-u>y\ \Bigg|\ Y_{0,1}>u\right) =  H(\underline{\boldsymbol{z}}) \exp(-y).
\end{equation}
Limit~(\ref{paper4::htlimiteq2}) in turn has the interpretation that as $u$ tends to infinity, $(\underline{\bold{Y}} -\underline{\boldsymbol{\a}} Y_{0,1})Y_{0,1}^{-\underline{\boldsymbol{\b}}}$ and $(Y_{0,1}-u)$ are independent conditional on $Y_{0,1}>u$, and are distributed as $H$ and a standard exponential, respectively. 

In practice, we exploit these results by assuming they hold exactly above some high finite threshold $u>0$. So, we approximate the conditional distribution of $\underline{\bold{Y}} |Y_{0,1}=y$ for $y>u$, $\underline{\bold{y}} \in \R^{(2k-1)d-1}$ as
\begin{equation}\label{paper4::HT_formulation}
\P(\underline{\bold{Y}}\leq \underline{\bold{y}}\mid Y_{0,1}=y) = H\left(\frac{\underline{\bold{y}} - \underline{\boldsymbol{\a}} y}{y^{\underline{\boldsymbol{\b}}}}\right),
\end{equation}
and we assume independence of $(\underline{\bold{Y}} -\underline{\boldsymbol{\a}} Y_{0,1})Y_{0,1}^{-\underline{\boldsymbol{\b}}}$ and $Y_{0,1}$. There is no finite-dimensional parametric form for $H$, so non-parametric methods are typically applied. However, we remark that there are applications of the conditional extreme value model where the copula $H$ is assumed to be Gaussian \citep{towe2019} or a Bayesian semi-parametric model is used \citep{lugrin2016bayesian}. For inference, see Section~\ref{paper4::sec::inference}.

\subsection{Multivariate Markov extremal model}\label{paper4::sec::model2}

For ease of presentation, we present the multivariate Markov extremal model (MMEM) of order $k$ only for a two-dimensional time-series $(\bold{Y}_t)_{t\in\Z}$ such that $\bold{Y}_t=(Y_{t,1},Y_{t,2})$ in the notation of Section~\ref{paper4::sec::myintro}, i.e., $\bold{Y}_t$ has standard Laplace margins. 
We only describe a forward model that is applicable to the post-peak period $\PPP^{\mathrm{post}}$, since the backward model has a similar construction. As mentioned in Section~\ref{paper4::sec::modellingstrat}, we apply a different forward MMEM model to $(\bold{Y}_{-t})_{t\in\Z}$ to yield the backward model for the pre-peak period $\PPP^{\mathrm{pre}}$. Concisely put, the MMEM exploits the HT model to estimate the distribution for $\bold{Y}_{t+k}$ conditional on $(\bold{Y}_{t},\dots,\bold{Y}_{t+k-1})$ when $Y_{t,1}>u$ for a large threshold $u>0$. 
As in Section~\ref{paper4::sec::model0}, for each $t\in\Z$, we define $\tilde{\bold{x}}_t\in\R^k\times\R^{k+1}$ to be an irregular matrix with $k+1$ rows and $2$ columns without the element that is on the first row and first column: 
\[
\tilde{\boldsymbol{x}}_t = \begin{pmatrix}  & x_{t,2} \\ x_{t+1,1} & x_{t+1,2} \\ \vdots & \vdots \\ x_{t+k,1} & x_{t+k,2} \end{pmatrix}. 
\]
Then, we assume that for a large threshold $u>0$, there exist parameters $\tilde{\boldsymbol{\a}}_0\in[-1,1]^{k}\times[-1,1]^{k+1}$, $\tilde{\boldsymbol{\b}}_{0}\in(-\infty,1)^{k}\times(-\infty,1)^{k+1}$, and a residual random variable $\tilde{\boldsymbol{\varepsilon}}_t$ on $\R^{k}\times\R^{k+1}$ with non-degenerate marginals 
 such that for $t\in\Z$
\[
\tilde{\bold{Y}}_t | (Y_{t,1}>u) = \tilde{\boldsymbol{\a}}_0 Y_{t,1} + Y_{t,1}^{\tilde{\boldsymbol{\b}}_0} \tilde{\boldsymbol{\varepsilon}}_t.
\]
Similar to \citet{winter2017}, for $t\in\Z$, $j\geq 1$ when $Y_{t+j,1}>u$, we then get
\[
[Y_{t+k+j,1}\ Y_{t+k+j,2}] | (\bold{Y}_{t+j:t+k+j-1},Y_{t+j,1}>u)= [\a_{k,1},\ \a_{k,2}] Y_{t+j,1} + Y_{t+j,1}^{[\b_{k,1},\ \b_{k,2}]} \cdot \boldsymbol{\varepsilon}^{C}_{k,1:2},
\]
where $\boldsymbol{\varepsilon}^{C}_{k,1:2}$ is short-hand notation for $[\boldsymbol{\e}_{k,1},\boldsymbol{\e}_{k,2}]$ conditional on $(\boldsymbol{\e}_{1:k-1,1},\boldsymbol{\e}_{0:k-1,2})$. For inference, we refer to Section~\ref{paper4::sec::inference}. 



\subsection{Extremal vector autoregression}\label{paper4::sec::model3}
Here, we introduce extremal vector autoregression (EVAR) for extremes of the process $(\bold{Y}_t)_{t\geq1}$. This model combines the HT model with a vector autoregressive model for the joint evolution of the time-series at high levels. Here we focus on the post-peak period, but note that the pre-peak period is modelled analogously. We define an EVAR model of order $k$ with parameters $\Phi^{(i)}\in \R^{d}\times \R^d$ for $i=1,\dots,k$ and $\bold{B}\in (-\infty,1)^d$ as
\begin{equation}\label{paper4::eqn::evar_model_formulation}
\bold{Y}_{t+k}|(\bold{Y}_{t},\dots,\bold{Y}_{t+k-1}) = \sum_{i=1}^{k} \Phi^{(i)} \bold{Y}_{t+k-i} + y^{\bold{B}}\boldsymbol{\varepsilon}_t,
\end{equation}
with $Y_{t,1}=y$ for $y>u$, where $u>0$ is a large threshold and $\boldsymbol{\varepsilon}_t$ is a $d$-dimensional multivariate random variable that has non-degenerate margins and is independent of $(\bold{Y}_{t},\dots,\bold{Y}_{t+k-1})$. Usually for a vector autoregressive model, parameter constraints would be imposed so that the resulting process is stationary. In the current extreme value context, stationarity is not of concern to us, since we reject trajectories that exceed the excursion maximum, and stop the process once the first component dips below threshold $u$. We define EVAR$_0$ as a special case of EVAR corresponding to $\bold{B}=\bold{0}$. EVAR$_0$ therefore has clear similarities with a regular vector autoregressive model \citep{tiao1981}, yet we emphasise that there is considerable difference between the two, since the parameters of EVAR$_0$ do not need to yield a stationary process, and the parameters of EVAR$_0$ are estimated using only extreme observations. To estimate the EVAR model, we adopt the same approach as that used to estimate the HT model, see Section~\ref{paper4::sec::inference}. As explained in Appendix~\ref{paper4::app::reparam}, the resulting parameter estimators $\hat{\Phi}^{(i)}$ are highly correlated. Hence a reparameterisation is introduced to reduce this correlation, and improve inference efficiency and computation.

For practical applications, an advantage of EVAR over MMEM is that it provides a lower-dimensional residual distribution when $k>1$ (with dimensions $d$ and $kd$, respectively). As a consequence, the EVAR residual distribution is less affected by the curse of dimensionality. A drawback of EVAR is that it might be insufficiently flexible to describe complex dependence well.

\subsection{Inference for conditional models}\label{paper4::sec::inference}
We discuss inference for each of the conditional extremes, MMEM and EVAR models with parameter vector $\boldsymbol{\theta}$. We discuss these together because they can be summarized in the same form. Specifically, let $\bold{W}=(W_1,\dots,W_d)$ be a $d$-dimensional random variable and assume that for some high threshold $u>0$,
\begin{equation}\label{paper4::eqn::generalmodel}
\bold{W}_{2:d}|(W_1>u) = \bold{g}_1(W_1;\boldsymbol{\theta}) + \bold{g}_2(W_1;\boldsymbol{\theta}) \boldsymbol{\e}
\end{equation}
for some parametric functions $\bold{g}_1(\,\cdot\,; \boldsymbol{\theta}):\R\to\R^{d-1}$ and $\bold{g}_2(\,\cdot\,;\boldsymbol{\theta}):\R\to\R^{d-1}_{>0}$, where 
\[
\bold{g}_1(x,\boldsymbol{\theta}):=(g_{1,2}(x,\boldsymbol{\theta}),\dots,g_{1,d}(x,\boldsymbol{\theta})),\ \text{and}\ \bold{g}_2(x,\boldsymbol{\theta}):=(g_{2,2}(x,\boldsymbol{\theta}),\dots,g_{2,d}(x,\boldsymbol{\theta})),\ \text{for}\ x\in\R
\]
where $\boldsymbol{\e}=(\e_2,\dots,\e_d)$ is a $(d-1)$-dimensional multivariate random variable that is non-degenerate in each margin and independent of $W_1$. As an example, for MMEM, $g_{1,j}(x)=\a_j x$ for some $\a_j$ and $g_{2,j}(x) = x^{\b_j}$ for some $\b_j$.

Next, assume that we have $n$ observations $\DDD:=\{\bold{w}_1,\dots,\bold{w}_n\}$ of the conditional random variable $\bold{W}|W_1>u$, where $\bold{w}_i=(w_{i1},\dots,w_{id})$ with $w_{i1}>u$ for $i=1,\dots,n$. We then infer $\boldsymbol{\theta}$ by calculating the likelihood of model~(\ref{paper4::eqn::generalmodel}) by temporarily assuming that the $\boldsymbol{\e}$ has a multivariate normal distribution with unknown mean $\boldsymbol{\mu}=(\mu_2,\dots,\mu_{d})$ and unknown diagonal covariance matrix $\Sigma=\boldsymbol{\s}^2I$ where $\boldsymbol{\s}^2 = (\s_2^2,\dots,\s^2_{d})$. These assumptions imply that the mean and the variance of $\boldsymbol{\e}$ are estimated simultaneously with the model parameters. The likelihood is then evaluated as
\[
L(\boldsymbol{\theta},\boldsymbol{\mu},\boldsymbol{\sigma}^2;\DDD) = \prod_{i=1}^{n} \prod_{j=2}^{d} \frac{1}{\sqrt{2\pi} \s_j g_{2,j}(w_{i1};\boldsymbol{\theta})} \exp\left\{-\frac{1}{2 \s_j^2}\left(\frac{w_{ij} - g_{1,j}(w_{i1}) - \mu_j g_{2,j}(w_{i1};\boldsymbol{\theta})}{g_{2,j}(w_{i1};\boldsymbol{\theta}) }\right)^2\right\}.
\]
Finally, the parametric assumption on the distribution of $\boldsymbol{\e}$ is discarded and estimated conditional on 
the parametric estimate $\hat{\boldsymbol{\th}}$ for $\boldsymbol{\th}$, with a kernel density $\hat{h}_{2:d}$ using the `observations' $\{\boldsymbol{\e}_i:\ i=1,\dots,n\}$ where $\boldsymbol{\e}_i=(\e_{i2},\dots,\e_{id})$ and 
\[
\e_{ij} := \frac{w_{ij} - \hat{g}_{1,j}(w_{i1};\hat{\boldsymbol{\th}})}{\hat{g}_{2,j}(w_{i1};\hat{\boldsymbol{\th}})}
\]
for $i=1,\dots,n$, $j=2,\dots,d$. In case of MMEM, we additionally require estimates for the density of a conditional random variable $\boldsymbol{\e}_{l+1:d|2:l} = (\e_{l+1},\dots,\e_{d})|(\e_2,\dots,\e_l)$ for some $l\in\{2,\dots,d-1\}$. Given the same set of observations, we estimate its conditional density $h_{l+1:d|2:l}$ as
\[
\hat{h}_{l+1:d|2:l}(\e_{l+1},\dots,\e_{d}|\e_2,\dots,\e_l) = \frac{ \hat{h}_{2:d}(\e_2,\dots,\e_d) }{\hat{h}_{2:l}(\e_2,\dots,\e_l) },
\]
where $h_{2:l}$ is estimated as the $(l-1)$-dimensional marginal of $\hat{h}_{2:d}$.

\section{Case Study - Northern North Sea}\label{paper4::sec::casestudy}

\subsection{Overview}

We apply MMEM, EVAR and a historical matching procedure (introduced in Section~\ref{paper4::sec::model1}, henceforth referred to as HM) to characterise excursions of significant wave height $H_S$ and wind speed $W_s$ with directional covariates for a location in the northern North Sea. Our goal is to estimate parsimonious predictive models for the joint evolution of $H_S$ and $W_s$ time-series conditional on $H_S$ being large.

In Section~\ref{paper4::sec::CSdata}, we describe the available met-ocean data. In Section~\ref{paper4::sec::drcmodel}, we outline a model for the evolution of storm direction that is needed for our time-series models. Section~\ref{paper4::sec::model1} then summarises the HM procedure, and in Section~\ref{paper4::sec::response}, we introduce structure variable responses that approximate fluid drag loading on a marine structure such as a wind turbine or coastal defence. Finally, in Section~\ref{paper4::sec::simulationstudy}, we compare the predictive performance of MMEM and EVAR (over a set of model orders) with the HM method in estimating structure variables for withheld intervals of time-series. 


\subsection{Data}\label{paper4::sec::CSdata}

We have $53$ years of hindcast data
\[
\DDD:=\{(H_{S,i},W_{s,i},\theta^H_i,\theta^W_i):\ i\in\TTT\}
\]
indexed with finite $\TTT\subset \Z_{\geq 1}$ consisting of time-series for four three-hourly met-ocean summary statistics at a location in the northern North Sea \citep{WAM09}: significant wave height ($H_{S,i}$ in metres), wind speed ($W_{s,i}$ in metres per second), wave direction ($\theta^H_i$ in degrees) and wind direction ($\theta^W_i$ in degrees) for each $i\in\TTT$. To use MMEM and EVAR, we transform significant wave height and wind speed onto Laplace marginals: $H_{S,i}| \theta_i^H \mapsto H^{\mathrm{L}}_{S,i}$ and $W_{s,i} |\theta_i^W \mapsto W^{\mathrm{L}}_{s,i}$, e.g., using directional marginal extreme value models for the tails \citep{chavez2005}, but ignoring seasonality. This part of the analysis has been reported on numerous occasions, see for example \citet{randell2015}. Because the marginal transformation includes direction as a covariate and because direction is not constant during an excursion, we also establish a model for the directional evolution of excursions in order to transform them between standard and original margins, see Section~\ref{paper4::sec::drcmodel}.

Let $D^L$ be the collection of the transformed data
\[
\DDD^L:=\{(H_{S,i}^L,W_{s,i}^L,\theta_i^H,\theta_i^W):\ i\in\TTT\}.
\] 
To define excursions in $\DDD^L$, we set the excursion threshold $u$ equal to the $95\%$ percentile of a standard Laplace distribution, i.e., $u\approx 2.3$, yielding $1,467$ observations of extreme excursions $\EEE_u$. This choice of threshold is not unusual as similar conclusions are drawn for excursion thresholds that are slightly different from our original choice. 

\begin{figure}[!htbp]
\centering
\includegraphics[width=\textwidth]{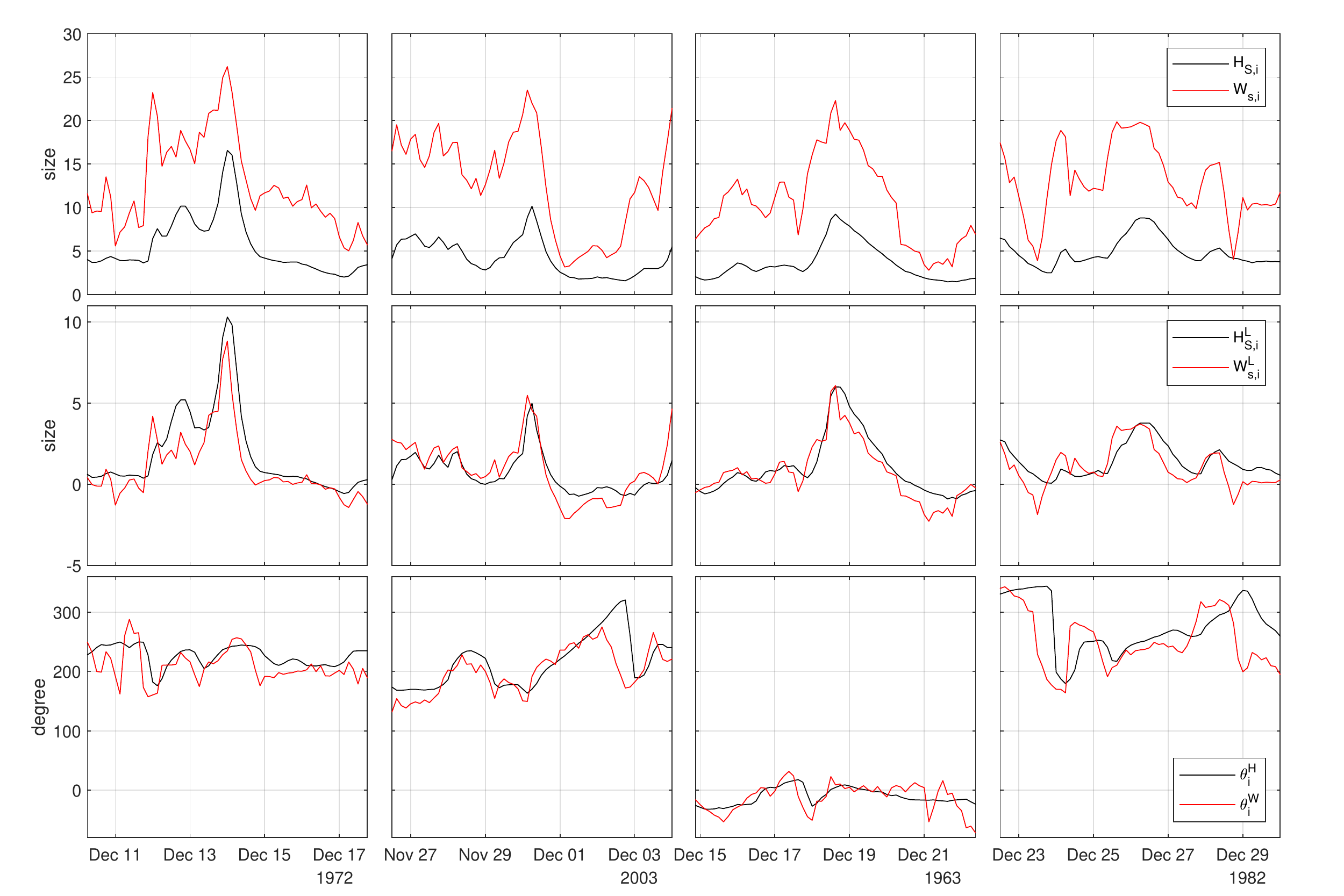}
\caption{Intervals of oceanographic time-series: (top) key variables: significant wave height $H_{S,i}$ and wind speed $W_{s,i}$ on original margins; (middle) on Laplace margins; (bottom) covariates: wave direction $\theta_i^H$ and wind direction $\theta_i^W$. The four columns correspond to time periods that contain the $100\%$, $95\%$, $90\%$ and $85\%$ empirical percentiles of $H_{S,i}$, respectively.}
\label{paper4::fig::dataslice}
\end{figure}

Figure~\ref{paper4::fig::dataslice} shows four intervals of the time-series chosen to contain the observations corresponding to the $100\%$, $95\%$, $90\%$ and $85\%$ sample percentiles of the set of excursion maximum significant wave heights, on original and standard Laplace margins, with directional covariates. Excursions are centred around extreme events. There is a large dependence of $H_S$ and $W_s$ on both original and standard margins. Moreover, variables associated to significant wave height, i.e., $H_S$, $H_S^L$ and $\th^H$, are much smoother than their wind speed counterparts. Additionally, the directional covariates $\th^H$ and $\th^W$ centre around each other with no large deviations during extreme events.

\begin{figure}[!htbp]
\centering
\includegraphics[width=\textwidth]{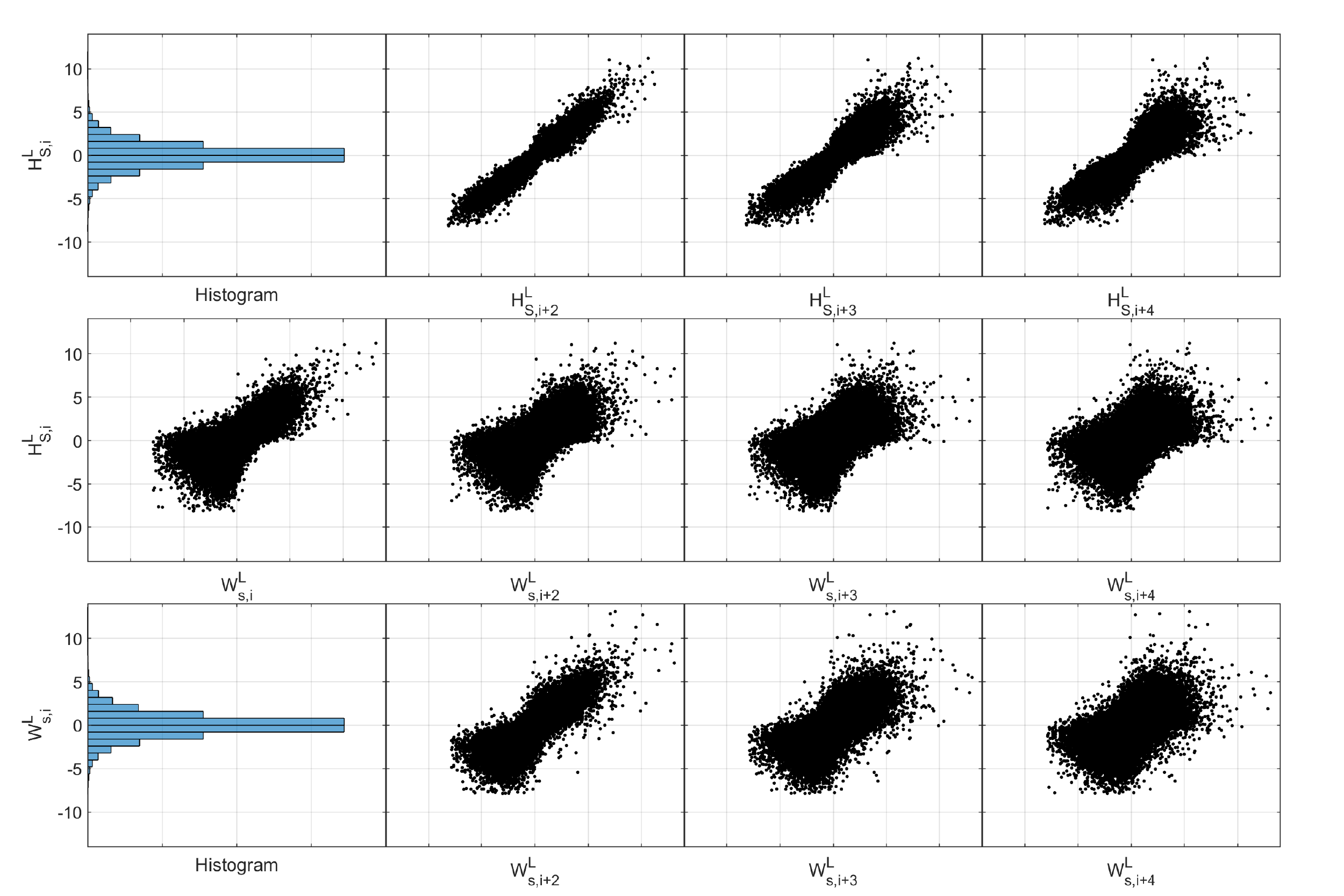}
\caption{Matrix plot of observed $H^{\mathrm{L}}_{S,i}$ and $W^{\mathrm{L}}_{s,i}$ at various time lags up to lag $4$ (corresponding to $12$ hours in real time) including cross dependece.}
\label{paper4::fig::matrixplot}
\end{figure}

In Figure~\ref{paper4::fig::matrixplot}, we visualize the (across variable joint) dependence of key variables $H^L_S$ and $W^L_s$ on Laplace scale at time lags up to lag $4$ using a series of scatterplots where a unit of lag corresponds to three hours of observation time. The figure illustrates the complex dependence of the bivariate time-series of significant wave height and wind speed on Laplace margins. As expected, we observe (slow) convergence to an independent variable model as lag increases. Most notably, we observe a similar level of dependence of $(H_{S,t}^L,W_{s,t+4}^L)$ and $(W_{s,t}^L,W_{s,t+4}^L)$ which suggests counter-intuitively that $H^L_{S,t}$ would be a better predictor for $W^L_{s,t+4}$ than $W_{s,t}^L$.

\begin{figure}[!htbp]
\centering
\includegraphics[width=\textwidth]{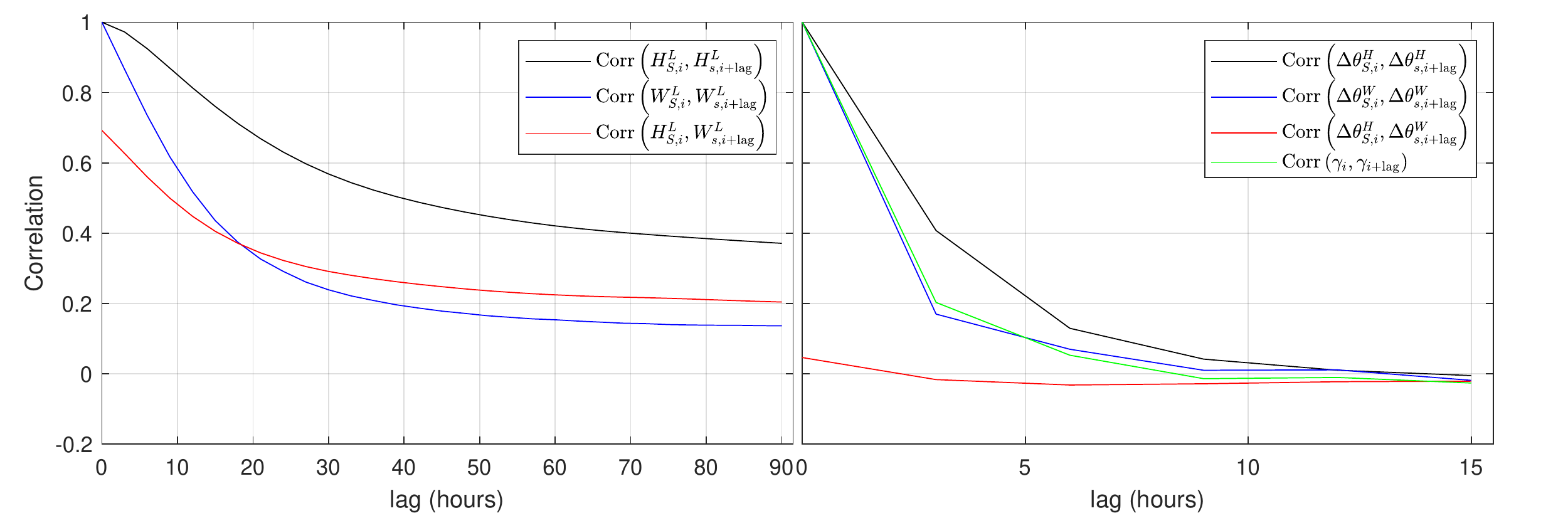}
\caption{Estimated correlation and cross-correlation at various time lags of: (left) the key variables on Laplace margins: $H_{S,i}^L$ and $W_{s,i}^L$; (right) the covariates: change in wave direction $\Delta \theta_i^H := (\theta_{i+1}^H - \theta_i^H,\ \mathrm{mod}\ 360)$, change in wind direction $\Delta \theta_i^W:= (\theta_{i+1}^W - \theta_i^W,\ \mathrm{mod}\ 360)$ and $\gamma_i$, see definition~(\ref{paper4::def::gammat}).}
\label{paper4::fig::correlation}
\end{figure}

In Figure~\ref{paper4::fig::correlation}, we plot (cross) correlation functions for these variables, and also for the change in directional covariates at various lags. These show that the dependence of $(H_{S,t}^L,H_{S,t+\tau}^L)$ decays relatively slowly as $\tau$ grows to $90$ hours, and that indeed the cross dependence between $(H_{S,t}^L,W_{s,t+\tau}^L)$ is larger than the dependence of $(W_{s,t}^L,W_{s,t+\tau}^L)$ for large $\tau$. Finally, the correlation plot of the change in directional covariates $\Delta \theta_{S,i}^H:=(\theta_{S,i+1}^H - \theta_{S,i}^H,\ \mathrm{mod}\ 360)$ and $\Delta\theta_{s,i}^W:=(\theta_{s,i+1}^W-\theta_{s,i}^W,\ \mathrm{mod}\ 360)$ on the right shows that a first order model for these covariates is appropriate since the correlations nearly vanish at lag $2$ (for wind and wave) or $6$ hours (for all other combinations).



\subsection{Directional model}\label{paper4::sec::drcmodel}

We model wave direction $\theta_i^H$ in a similar fashion as~\citet{tendijck2019}, summarised as follows. Let $\III\subset\TTT$ be the set of indices of the original data that correspond to all observations of any excursion. Next, let $\{d({\theta}_{i+1}^H,{\theta}^H_{i}):\ i \in \III\}$ be the set of changes in wave directions, where $d(\th,\th')=(\th-\th'+180;\ \mod 360)-180\in[-180,180)$ denotes the circular difference of $\th$ and $\th'$ in degrees. In our application, the set of changes in wave directions during excursions do not contain values close to $-180$ or $180$. In particular, all of the observed changes centre around $0$.

 For $i\in\III$, we transform observations $d({\theta}_{i+1}^H,{\theta}^H_{i})\mapsto \delta^H_i :=\Phi^{-1}(\hat{F}(d({\theta}_{i+1}^H,{\theta}^H_{i})))$ on Gaussian margins,  where $\hat{F}$ denotes the empirical distribution function of the set of changes in wave directions. 
Assume that $\{\d_i^H:\ i\in\III\}$ are realisations of the random variables $\{\Delta^H_i:\ i\in\III\}$. 
We estimate the following autoregressive model for $\Delta^H_t$ of order $p_1=1,2,3,\dots$ with parameters $\phi^{\mathrm{H}}_{j}\in\R$ for $j=1,\dots,p_1$ as
\begin{equation}\label{paper4::model_Wavedrc}
\Delta^H_{t} | (\Delta^H_{t-1},\dots,\Delta^H_{t-p_1}) = \sum_{j=1}^{p_1} \phi^{\mathrm{H}}_{j} \Delta^H_{t-j} +  \zeta(H_{S,t}) \e_t,
\end{equation}
 where $\e_t$ is a standard Gaussian random variable, and standard error $\zeta(h)$ is given by
\[
\zeta^2(h) = \lambda_{1} + \lambda_{2} \exp(-\lambda_{3} h)
\]
with $\lambda_{j'}>0$ for $j'=1,2,3$, see~\citet{tendijck2019}. In particular, the standard error $\zeta(h)$ decays as $h$ grows due to the significantly larger amounts of energy needed to change the direction of more severe sea states. The parameters of this model are inferred with maximum likelihood, and in contrast to the inference discussed in Section~\ref{paper4::sec::inference}, we do not reject the assumption that $\e_t$ is a standard Gaussian. In practice, we use $p_1=1$ in line with~\citet{tendijck2019}.

Given model~(\ref{paper4::model_Wavedrc}), we propose the following model
\begin{equation}\label{paper4::def::gammat}
\theta_t^W = \theta_t^H + \g_t\ \ \ \mathrm{mod}\ 360
\end{equation} 
for wind direction $\theta_t^W$ conditional on wave direction $\theta_t^H$, where $\g_t$ is a zero-mean stationary AR($p_2$) process. That is, there exist parameters $\phi^{\mathrm{W}}_j\in\R$, $1\leq j\leq p_2$, and a non-degenerate residual distribution $r_t$ independent of $\g_{t-j}$ for $j\geq1$, such that
\[
\g_{t}|(\g_{t-1},\dots,\g_{t-p_2}) = \sum_{j=1}^{p_2} \phi^{\mathrm{W}}_j \g_{t-j} + r_{t},
\] 
and such that the polynomial $1-\sum_{j=1}^{p_2} \phi^{\mathrm{W}}_j z^j$ has roots outside the unit circle. The model parameters and the distribution of $r_t$ are inferred as described in Section~\ref{paper4::sec::inference} conditional on the model order $p_2$, which is selected by investigating the correlation function in Figure~\ref{paper4::fig::correlation} and the partial autocorrelation function of $\g_t$ (not reported). In our application, we conclude that $p_2=1$ is sufficient.


\subsection{Historical matching}\label{paper4::sec::model1}

An empirical method for simulating excursions is described in \citet{feld2015} and termed historical matching (HM) in this work. They model trajectories of significant wave height, wave direction, season and wave period during extreme events. The key assumption they make is that storm trajectory (or excursion) profiles are not independent of storm maximum conditions.  
Specifically, the HM approach is a composition of four models:
(i) a model for storm maximum wave direction; 
(ii) a model for storm maximum significant wave height conditional on storm maximum wave direction; 
(iii) a model that selects at random a historical storm trajectory with similar storm maximum characteristics to that simulated;
(iv) a model that adjusts the historical storm trajectory by matching storm maximum characteristics of simulated and historical storms.

Specific details of the individual models are as follows, but this level of detail is not required for understanding the impact of the core methodology developments in Section~\ref{paper4::sec::casestudy}. For model (i), we simply sample at random from the observed wave directions associated with storm maximum significant wave height (excursion maximum). In model (ii), storm maximum significant wave height are modelled using a generalised Pareto distribution conditional on the sampled storm maximum wave direction using a generalised additive model with the parameters as B-splines conditional on directional covariates \citep{chavez2005}.  In model (iii), we use a distance measure to calculate the dissimilarity between pairs of storm maximum significant wave heights and storm maximum wave directions for simulated and historical trajectories. Here, we use the heuristic recommended by \citet{feld2015} ensuring that a difference of $5$ degrees in storm maximum wave direction corresponds to the same dissimilarity as $0.5m$ of difference in storm maximum significant wave height; one of the closest $20$ matching storms is then selected at random for associated with the simulated storm maximum. In model (iv), we match the variables of the chosen historical trajectory as follows:
(a) the historical significant wave height series are multiplied by the ratio of the simulated maximum significant wave height to the maximum of the historical significant wave height; 
(b) the historical wave directions are shifted such that the storm maximum wave directions of simulated and historical trajectories coincide; 
(c) the associated historical wind directions are rotated in the exact same way as wave direction; 
(d) for the full set of historical storm maxima, storm maximum associated wind speed $W_s^{M}$ (namely the value of wind speed at the time point corresponding to the storm maximum event) conditional on storm maximum significant wave height $H_S^{M}$ is described using linear regression with parameters $\b_0,\b_1\in\R$, $\s>0$:
\[
W_s^{M}| H_S^M = \b_0 + \b_1 H_S^{M} + \s \e
\]
with $\e$ a standard normal random variable;
(e) wind speed for the selected historical trajectory is scaled linearly such that it agrees with the storm maximum associated wind speed from (d).

 Perhaps the main deficiencies of the HM approach are (i) it does not provide a means for modelling the extremal temporal dependence characteristics of excursions, and the extremal dependence between different components of the time-series for excursions to levels beyond those observed in the historical sample, and (ii) it does not provide a model framework for the assessment of fit or uncertainty propagation.

\subsection{Response variable}\label{paper4::sec::response}
To measure the practical impact of extreme met-ocean excursions, we define structure response variables for a simple hypothetical marine offshore facility. A structure response variable is a function of the met-ocean variables, key to assessing the integrity of the design of a physical structure of interest. Specifically, we consider a structure in the form of a unit cube standing above the water, supported by thin rigid legs, with vertical cube faces aligned with cardinal directions. Only wave and wind impact on the cube itself is of interest to us, and we neglect the effects of other oceanic phenomena such as swell, surge, tide, and potential climate non-stationarity. For simplicity, we also assume that when $H_S<h$, for some known value $h>0$, the wave impact on the structure is negligible, and structural response is dominated by wind. When $H_S\geq h$, we assume that wave impact increases cubically with $H_S$ and quadratically with $W_s$ (see \citealt{morison1950force} and \citealt{ma2020effective} for supporting literature). Hence, the impact of an extreme excursion on the structure is defined by the instantaneous response variable $R$
\[
R(H_{S},W_{s},\theta^H,\theta^W; c,h) =  \begin{cases} c\cdot I^2_{W}(W_s,\th^H-\th^W) & \text{for}\ H_{S} < h, \\  c\cdot I^2_{W}(W_s,\th^H-\th^W) + A(\th^H)\cdot(H_{S} - h)\cdot H_{S}^2 & \text{for}\ H_{S} \geq h, \end{cases}
\]
where $I_W:\R_{>0}\times[-180,180)\to\R$ is the inline wind-speed, defined below, $A:[-180,180)\to[1,\sqrt{2}]$ is the exposed cross-sectional area of the cube, see below, and the parameter $c>0$ is specified such that both significant wave height and wind speed have an approximately equal contribution to the largest values of $R$. Here both $c$ and $h$ are values that can be changed by altering structural features. The exposed cross-sectional area $A(\theta)\in[1,\sqrt{2}]$ of the cube is given by
\[
A(\theta^H) :=  1/\cos([(\theta^H+45;\ \mathrm{mod} 90)-45]\cdot \pi/180)
\]
for a given wave direction $\theta^H$. The inline wind-speed $I_{W}$ is the component of the wind speed in the direction of the wave given by
\[
I_{W}(W_s,\theta^H-\theta^W) = W_{s} \cos((\theta^H - \theta^W)\cdot \pi/180).
\]
To simplify notation, we write $R_i(c,h):=R(H_{S,i},W_{s,i},\theta_i^H,\theta_i^W; c,h)$ for $i\in\TTT$. To define a structure response for a complete excursion $\EEE_u$, we write
\[
\EEE_u:=\{(H_{S,i},W_{s,i},\Theta_i^H,\Theta_i^W):\ a \leq i \leq b\},
\]
for some $a<b$ such that for a threshold $u>0$ (on Laplace margins) $H^L_{S,i}>u$ for $a\leq i\leq b$ and $H^L_{S,a-1},H^L_{S,b+1}\leq u$. Next, let $i^*:=i^*(\EEE_u)$ be the time of the excursion maximum, i.e., $H_{S,i^*}$ is the maximum of $H_{S,i}$ over $\EEE_u$. We define two natural structure response variables representing the maximum impact of an excursion $\max_{\{a\leq i\leq b\}} R_i(c,h)$, and the cumulative impact of an excursion $\sum_{\{a\leq i\leq b\}} R_i(c,h)$, respectively. For our application, we consider slight alterations $R^{\max}(c,h,\EEE_u)$ and $R^{\mathrm{sum}}(c,h,\EEE_u)$
\[
R^{\max}(c,h,\EEE_u) := \max_{\{a\leq i\leq b,\ |i-i^*|>2\}} R_i(c,h),\ \ \ \ \ R^{\mathrm{sum}}(c,h,\EEE_u) := \sum_{\{a\leq i\leq b,\ |i-i^*|>2\}} R_i(c,h).
\]
That is, we consider responses that do not depend directly on the characteristics of the excursion near to the excursion maximum, to exaggerate the dependence of the structure variables on pre-peak and post-peak periods compared to the period of the peak, and hence the importance of estimating good models for the pre-peak and post-peak periods using MMEM or EVAR. Moreover, we define $R^{\max}(c,h)$ and $R^{\mathrm{sum}}(c,h)$ as the random structure responses related to a random excursion. 

\subsection{Model comparisons}\label{paper4::sec::simulationstudy}
Here, we use our time-series models to characterise extreme excursions for the met-ocean data $\DDD$ of Section~\ref{paper4::sec::CSdata} with structure responses $R^{\max}$ and $R^{\mathrm{sum}}$.  First, we investigate the model fits, then we describe our model comparison procedure, and finally we assess model performance using a visual diagnostic.

We fit EVAR, EVAR$_0$ and MMEM with model orders $k=1,2,\dots,6$ to data $\DDD^L$. The fitting of these 18 models is a two-stage procedure. In the first part, we fit (six) conditional extremes models for the period of the peak $\PPP^k_0$ for each $k$. In the second part, we fit $2\cdot 18=36$ models to the pre-peak $\PPP^{\mathrm{pre}}$ and post-peak $\PPP^{\mathrm{post}}$ periods. In Table~\ref{paper4::table::prm_est0}, we report parameter estimates of the period of the peak model, and in Tables~\ref{paper4::table::prm_est1}-\ref{paper4::table::prm_est1_pre}, we report parameter estimates of MMEM on $\PPP^{\mathrm{post}}$ and $\PPP^{\mathrm{pre}}$, respectively. Finally, we report parameter estimates of EVAR on $\PPP^{\mathrm{post}}$ and $\PPP^{\mathrm{pre}}$ in Tables~\ref{paper4::table::prm_est2}-\ref{paper4::table::prm_est2_pre}, respectively. These indicate that all models agree on some level of asymptotic independence at each lag (coefficients of $\tilde{\boldsymbol{\a}}_0$ are less than $1$) with decreasing levels of dependence as lag increases, which can be seen by decreasing coefficients of $\tilde{a}_0$ for entries further down the table. We remark that for EVAR(2) on $\PPP^{\mathrm{pre}}$, the coefficient of $H_S$ at time $t+1$ $(0.96)$ is larger than the coefficient of $W_s$ at time $t+1$ $(0.50)$ for estimating $W_s$ at time $t+2$. This has the interpretation that significant wave height might be a better predictor for wind speed than wind speed itself, also suggested by Figure~\ref{paper4::fig::correlation}.

\begin{table}[!htbp]
\centering
\begin{tabular}{c c}
$\underline{\boldsymbol{\a}}$  &\\ \hline
0.54 (0.53,0.55) & 0.58 (0.56,0.59) \\
0.67 (0.66,0.68) & 0.75 (0.73,0.76) \\
0.86 (0.85,0.87) & 0.91 (0.90,0.93) \\
& 0.86 (0.84,0.87) \\
0.88 (0.87,0.88) & 0.61 (0.59,0.62) \\
0.73 (0.72,0.74) & 0.46 (0.45,0.47) \\
0.61 (0.60,0.62) & 0.36 (0.34,0.37) \\
\end{tabular}
\hspace{1cm}
\begin{tabular}{c c}
$\underline{\boldsymbol{\b}}$  &\\ \hline
0.68 (0.59,0.72) & 0.36 (0.31,0.52) \\
0.76 (0.66,0.82) & 0.34 (0.32,0.41) \\
0.82 (0.61,1.00) & 0.08 (0.07,0.14) \\
& 0.27 (0.20,0.29) \\
0.75 (0.52,0.96) & 0.46 (0.33,0.47) \\
0.64 (0.62,0.80) & 0.46 (0.35,0.56) \\
0.49 (0.48,0.66) & 0.16 (-0.03,0.35) \\
\end{tabular}
\caption{Estimates of model parameters $\underline{\boldsymbol{\a}}$ and $\underline{\boldsymbol{\b}}$ for the period of the peak $\PPP_0^k$ with model order $k= 4$. Also shown in parentheses are $90\%$ bootstrap confidence intervals. The structure of the irregular matrix estimates of $\underline{\boldsymbol{\a}}$ and $\underline{\boldsymbol{\b}}$ is explained in Section~\ref{paper4::sec::model0}.}
\label{paper4::table::prm_est0}
\end{table}

\begin{table}[!htbp]
\centering
\begin{tabular}{c c}
$\tilde{\boldsymbol{\a}}_{0}$  &\\ \hline
  & 0.74 (0.73, 0.75) \\
 0.86 (0.86, 0.87) & 0.56 (0.55, 0.57) \\
 0.73 (0.72, 0.74) & 0.44 (0.43, 0.45) \\
 0.63 (0.62, 0.64) & 0.35 (0.34, 0.37) \\
 0.55 (0.54, 0.56) & 0.29 (0.27, 0.31) \\
\end{tabular}
\hspace{1cm}
\begin{tabular}{c c}
$\tilde{\boldsymbol{\b}}_{0}$  &\\ \hline
 & 0.37 (0.29, 0.38) \\
 0.36 (0.35, 0.44) & 0.36 (0.26, 0.46) \\
 0.46 (0.45, 0.51) & 0.31 (0.20, 0.39) \\
 0.44 (0.43, 0.54) & 0.13 (0.01, 0.22) \\
 0.29 (0.28, 0.51) & 0.05 (-0.06, 0.16) \\
\end{tabular}
\caption{Estimates of MMEM model parameters $\tilde{\boldsymbol{\a}}_0$ and $\tilde{\boldsymbol{\b}}_0$ with model order $k= 4$ for $\PPP^{\mathrm{post}}$. Also shown in parentheses are $90\%$ bootstrap confidence intervals. The structure of the irregular matrix estimates of $\tilde{\boldsymbol{\a}}$ and $\tilde{\boldsymbol{\b}}$ is explained in Section~\ref{paper4::sec::model2}.}
\label{paper4::table::prm_est1}
\end{table}

\begin{table}[!htbp]
\centering
\begin{tabular}{c c}
$\tilde{\boldsymbol{\a}}_{0}$  &\\ \hline
  & 0.93 (0.92, 0.95) \\
 0.84 (0.83, 0.84) & 0.88 (0.87, 0.90) \\
 0.67 (0.67, 0.69) & 0.73 (0.72, 0.74) \\
 0.56 (0.55, 0.57) & 0.60 (0.59, 0.61) \\
 0.48 (0.47, 0.50) & 0.50 (0.49, 0.52) \\
\end{tabular}
\hspace{1cm}
\begin{tabular}{c c}
$\tilde{\boldsymbol{\b}}_{0}$  &\\ \hline
 & 0.06 (0.05, 0.09) \\
 0.29 (0.28, 0.49) & 0.10 (0.09, 0.16) \\
 0.46 (0.45, 0.55) & 0.25 (0.24, 0.36) \\
 0.52 (0.51, 0.59) & 0.32 (0.31, 0.45) \\
 0.42 (0.41, 0.56) & 0.37 (0.27, 0.44) \\
\end{tabular}
\caption{Estimates of MMEM model parameters $\tilde{\boldsymbol{\a}}_0$ and $\tilde{\boldsymbol{\b}}_0$ with model order $k= 4$ for $\PPP^{\mathrm{pre}}$.  Also shown in parentheses are $90\%$ bootstrap confidence intervals. The structure of the irregular matrix estimates of $\tilde{\boldsymbol{\a}}$ and $\tilde{\boldsymbol{\b}}$ is explained in Section~\ref{paper4::sec::model2}.}
\label{paper4::table::prm_est1_pre}
\end{table}

\begin{table}[!htbp]
\centering
\begin{tabular}{c | c c}
EVAR(1) \\ \hline
$\Phi^{(1)}$ &0.75 (0.70, 0.80) & 0.08 (0.07, 0.10) \\
 &0.10 (0.06, 0.14) & 0.65 (0.62, 0.68) \\
\hline
$\bold{B}$ &0.43 (0.37, 0.53) & 0.39 (0.29, 0.48) \\
\end{tabular}
\begin{tabular}{c | c c}
EVAR(2) \\ \hline
$\Phi^{(1)}$ &1.12 (1.08, 1.16) & 0.17 (0.15, 0.19) \\
 &0.29 (0.23, 0.34) & 0.80 (0.76, 0.84) \\
\hline
$\Phi^{(2)}$ &-0.32 (-0.36, -0.28) & -0.14 (-0.16, -0.13) \\
 &-0.21 (-0.26, -0.16) & -0.10 (-0.14, -0.07) \\
\hline
$\bold{B}$ &0.53 (0.52, 0.60) & 0.18 (0.08, 0.28) \\
\end{tabular}
\caption{Estimates of EVAR model parameters (Section~\ref{paper4::sec::model3}) with model order $k=1$ (left), $2$ (right) for $\PPP^{\mathrm{post}}$.  Also shown in parentheses are $90\%$ bootstrap confidence intervals.}
\label{paper4::table::prm_est2}
\end{table}

\begin{table}[!htbp]
\centering
\begin{tabular}{c | c c}
EVAR(1) \\ \hline
$\Phi^{(1)}$ &0.90 (0.84, 0.94) & -0.18 (-0.19, -0.16) \\
 &0.35 (0.31, 0.41) & 0.50 (0.46, 0.53) \\
\hline
$\bold{B}$ &0.40 (0.37, 0.63) & 0.15 (0.12, 0.24) \\
\end{tabular}
\begin{tabular}{c | c c}
EVAR(2) \\ \hline
$\Phi^{(1)}$ &1.31 (1.26, 1.35) & -0.19 (-0.21, -0.17) \\
 &0.96 (0.89, 1.03) & 0.50 (0.47, 0.54) \\
\hline
$\Phi^{(2)}$ &-0.46 (-0.53, -0.41) & 0.11 (0.09, 0.13) \\
 &-0.67 (-0.76, -0.59) & 0.10 (0.07, 0.14) \\
\hline
$\bold{B}$ &0.53 (0.51, 0.64) & 0.29 (0.26, 0.43) \\
\end{tabular}
\caption{Estimates of EVAR model parameters (Section~\ref{paper4::sec::model3}) with model order $k=1$ (left), $2$ (right) for $\PPP^{\mathrm{pre}}$.  Also shown in parentheses are $90\%$ bootstrap confidence intervals.}
\label{paper4::table::prm_est2_pre}
\end{table}

\begin{figure}[!htbp]
\centering
\includegraphics[width=\textwidth]{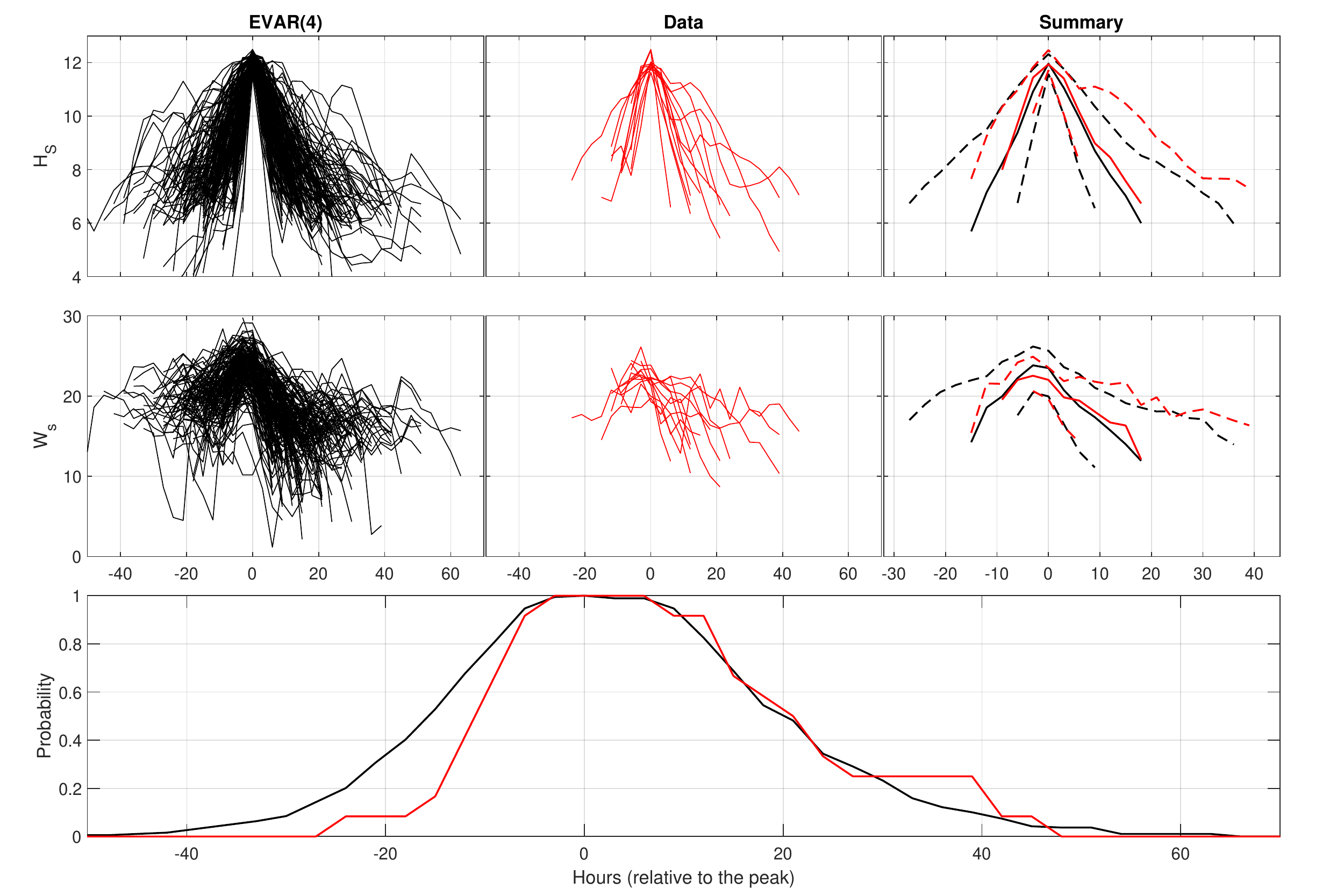}
\caption{Excursions of $H_S$ and $W_s$ from EVAR(4) model (left; black), and data (middle; right) on original margins such that storm peak significant wave height is in $[11.5,12.5]$; (right) summaries of the data (black) and EVAR(4) (red) excursions: median (solid), and the $10\%$ and $90\%$ quantiles (dashed). In the bottom panel, we plot survival probabilities for observed (black) and EVAR(4) (red) excursions relative to the time of the excursion maximum, see equation~(\ref{paper4::eqn::survprobofexcursion}).}
\label{paper4::fig::trajectories}
\end{figure}

For each of the $18$ models and HM, we simulate $20,000$ excursions to estimate model properties. First, we illustrate model characteristics for EVAR(4) in Figure~\ref{paper4::fig::trajectories} by plotting simulated excursions such that the excursion maximum significant wave height takes on values between $11.5$m and $12.5$m (left). We visually compare these with observed excursions for the same interval of excursion maxima (middle). On the right, we summarize simulated and observed excursions in terms of the median, the $10\%$ and $90\%$ percentiles of the sampling distribution at each time period. Finally, in the bottom panel we plot 
\begin{equation}\label{paper4::eqn::survprobofexcursion}
\P\left(\min\{ H_{S,i}^L:\ i=\min(0,\tau),\dots,\max(0,\tau)\} > u\ \Big|\ H_{S,0} \in [11.5,12.5]\right), 
\end{equation}
for $\tau\in\Z$, i.e., we plot the survival probability for an excursion relative to the time of the excursion maximum, conditional on the excursion maximum taking a value between $11.5$m and $12.5$m for both the simulated excursions and the observed excursions. We observe good agreement in the distribution of the length of an excursion with respect to the excursion maximum as both estimates are close to each other.

 In the supplementary material, we produce analogous plots for each of the 18 models considered and HM. We observe that EVAR(4) characterizes the period of the peak, and also the pre-peak and post-peak periods of the excursion well. Moreover, EVAR(4) also reproduces the observed excursion survival probability.

\begin{figure}[!htbp]
\centering
\includegraphics[width=\textwidth]{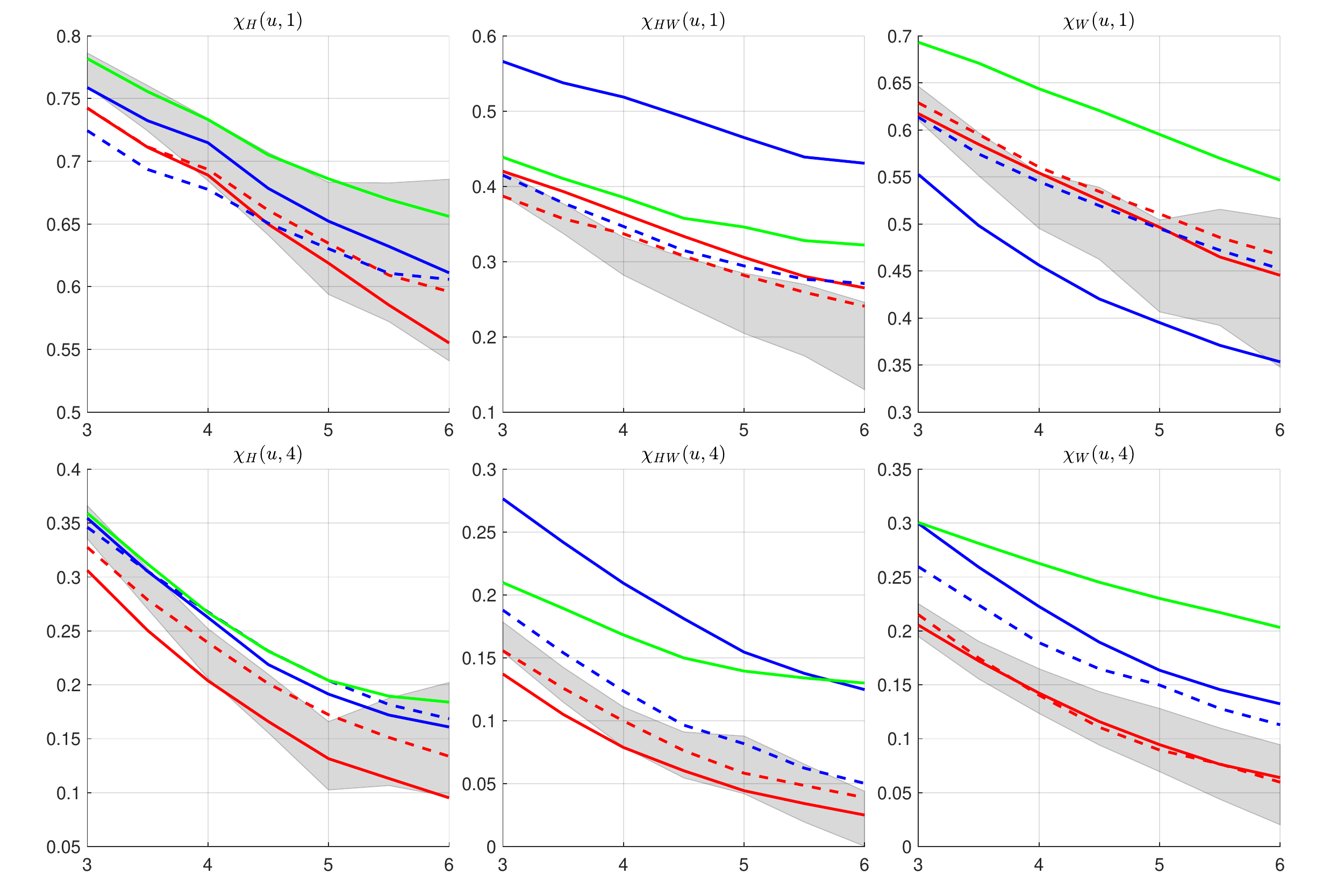}
\caption{Estimates of measures of extremal dependence across time lags $1$ and $4$, and variables given by $\chi_H$, $\chi_{HW}$ and $\chi_W$ (left, middle, and right respectively) for each of the models: EVAR (red), MMEM (blue), HM (green), data (grey). For EVAR and MMEM, we plot these estimates for different model orders $k=1$ and $k=4$ with line types: one (solid), four (dotted). Moreover, the grey region depicts the confidence bounds for empirical estimates of these extremal dependence measures from the data.}
\label{paper4::fig::modelfit}
\end{figure}

Next, in Figure~\ref{paper4::fig::modelfit}, we plot estimates of conditional probabilities $\chi_{H}(u,l):=\P(H_{S,t+l}^L>u\mid H_{S,t}^L>u)$, $\chi_{HW}(u,l):=\P(W_{s,t+l}^L>u\mid H_{S,t}^L>u)$, and $\chi_{W}(u,l):=\P(W_{S,t+l}^L>u\mid W_{S,t}^L>u)$ using EVAR, MMEM and HM with model orders $1$ and $4$, and we compare these with empirical estimates.\footnote{We leave out EVAR$_0$ in this analysis for conciseness since its estimates are very similar to the estimates obtained using EVAR of the same model order.} We make the following observations: HM is significantly worse at characterizing each of $\chi_H$, $\chi_W$ and $\chi_{HW}$ compared to EVAR and MMEM. Moreover, estimates obtained from EVAR of large enough order, e.g., $k\geq 4$, agree well with empirical estimates. MMEM, on the other hand, yields estimators that are slightly positively biased. In particular, larger model orders yield considerable improvements.

In Figure~\ref{paper4::fig::modelfit}, we discuss goodness-of-fit of each of the models. To compare MMEM and EVAR with each other and with HM, we take a similar approach to \citet{gandy2022}, who adjust standard cross-validation techniques to extreme value applications by taking a small training set and a larger test set. We select at random $25\%$ of the observed excursions for our training sample; the remaining $75\%$ forms our test sample. Below, we calculate performance statistics for the response variables by averaging over $50$ such random partitions of the sample.

For training, we fit EVAR, EVAR$_0$ and MMEM with model orders $k=1,2,\dots,6$ as explained in the second paragraph of this section. 
For each of the $18$ models and HM, we simulate $20,000$ excursions, calculate structure response variables $R^{\max}$ and $R^{\mathrm{sum}}$, and compare distributions of simulated structure response variables with those corresponding to the withheld test data. This is achieved by defining a dissimilarity distance function $D$ that measures the level of difference in tails of distribution functions. We select $20$ equidistant percentiles $p_1,\dots,p_{20}$ ranging from $97\%$ to $99.9\%$ corresponding to moderately extreme to very extreme levels with respect to the (smaller) training sample but not too extreme for the (larger) withheld data. We define the distance $D$ of distribution functions $F_M$ (of model $M$) and $F_E$ (an empirical distribution function) as the mean absolute relative error over these percentiles, i.e.,
\[
D(F_M,F_E;p_1,\dots,p_{20}) = \frac{1}{20}\sum_{j=1}^{20} \left|\frac{F_E^{-1}(p_j) - F_M^{-1}(p_j)}{F_E^{-1}(p_j)}\right|.
\]
We remark that in the above definition, we never divide by zero because we only use $D$ to measure the dissimilarity of distributions of positive random variables.

\begin{figure}[!htbp]
\centering
\includegraphics[width=0.75\textwidth]{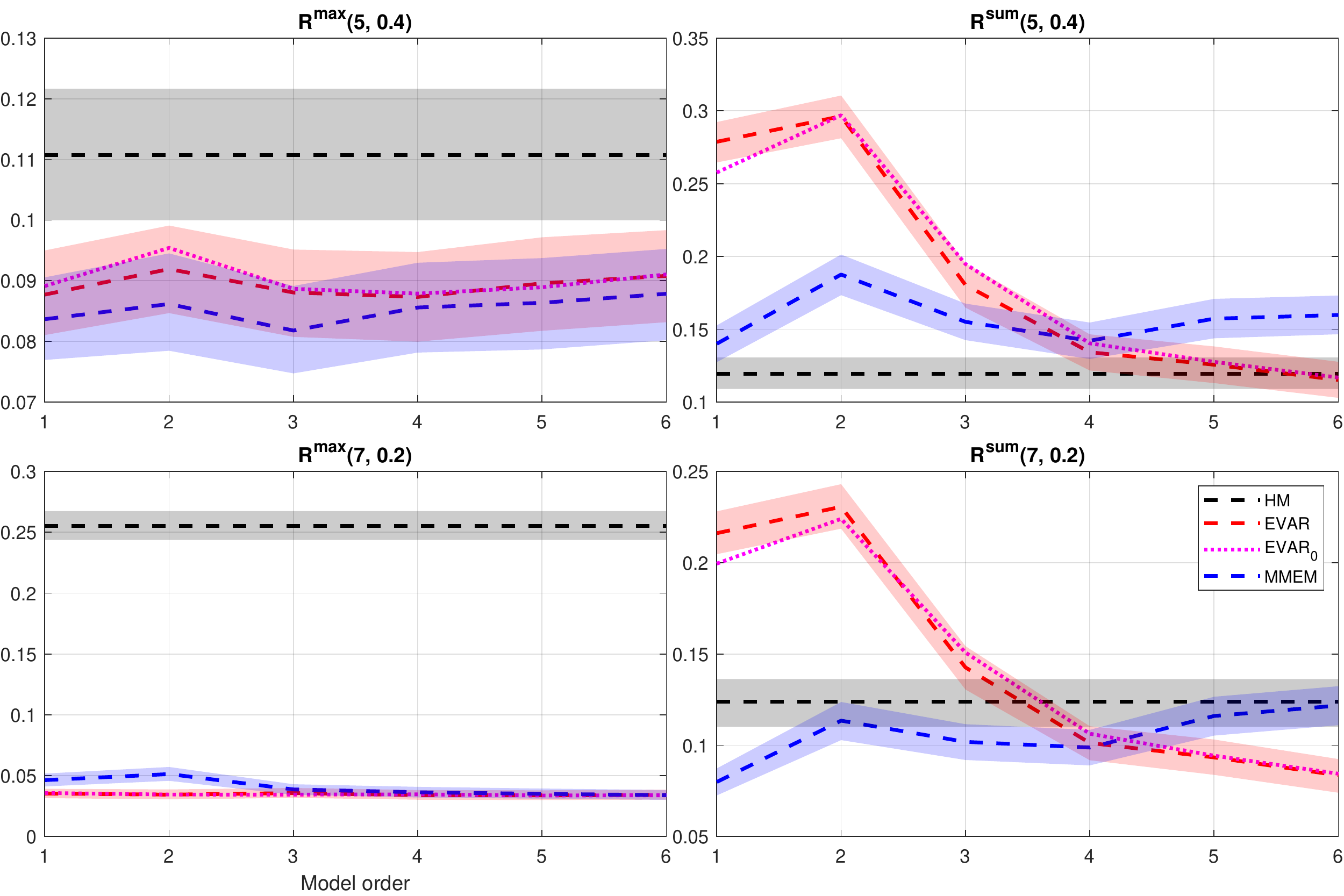}
\caption{Average mean relative errors of HM, EVAR, EVAR$_0$ and MMEM (dashed/dotted) and $80\%$ confidence regions (shaded) for estimating the distribution of structure responses using $25\%$ of data for training and $75\%$ of data for testing. For details, see the text.}
\label{paper4::fig::response}
\end{figure}

In Figure~\ref{paper4::fig::response}, we show the results for the $50$ random partitions of the original sample by plotting the average distance $D$ (with $80\%$ confidence intervals) for each model together with HM for four different structure response variables corresponding to two choices of $c$ and $h$ for each of $R^{\max}$ and $R^{\mathrm{sum}}$.  Note that similar studies for other values of $c$ and $h$ for $R^{\max}$ and $R^{\mathrm{sum}}$ were examined, and general findings are consistent with those illustrated in Figure~\ref{paper4::fig::response}. For legibility, we omit confidence bands for EVAR$_0$ since the difference with EVAR is minimal. Model selection now involves choosing the model that yields the smallest average dissimilarity $D$ whilst keeping the model order as low as possible.

We make a number of observations. For the $R^{\max}$ response, EVAR and MMEM clearly outperform HM regardless of model order. However, for the $R^{\mathrm{sum}}$ response, high order (e.g., $k=4,5,6$) EVAR and MMEM are necessary to be competitive with HM. We observe also that performance of EVAR and MMEM does not significantly improve or worsen for $k>4$. This finding is further supported with an unpublished study with Markov model orders of $k\leq10$. We note that llustrations of excursions in the supplementary material demonstrate that MMEM(1) does not explain the variability of the pre-peak and post-peak periods well. 

By looking at the average relative errors in $R^{\max}$ and $R^{\mathrm{sum}}$ of our proposed selection of methods, we conclude that a third or fourth order MMEM and a fourth order EVAR are competitive models within their class. Since these models have similar performance, we prefer EVAR(4) because of its simpler two-dimensional residual distribution. 



\section{Conclusions and discussion}
In this paper, we provide models for extreme excursions of multivariate time-series. Excursions are characterized by a three-stage modelling procedure for the period of the peak, the pre-peak and the post-peak periods. We model the period of the peak using the conditional extremes framework \citep{heffernan2004conditional}, and for the pre-peak and post-peak periods, we define two classes of time-series models: MMEM, motivated by the Markov extremal model of \citet{winter2017}; and EVAR, an extreme-value extension of a vector autoregressive model. We compare these excursion models with a baseline historical matching method, motivated by \citet{feld2015}. We find that the excursion models - for a reasonably informed choice of $k$, the order of the Markov process - are at least competitive with historical matching and often outperform it in the estimation of the tail of a range of notional structure response variables for a met-ocean application in the northern North Sea.

Statistical modelling of extreme excursions of multivariate time-series is difficult as it requires the estimation of complex model forms. MMEM requires the estimation of the conditional distribution of high-dimensional residual random variables and EVAR is highly parameterized. Nevertheless, for realistically sized directional samples of significant wave height and wind speed time-series, we found that MMEM(3), MMEM(4) and EVAR(4) perform well. Even when the empirical historical matching procedure is competitive, adoption of an excursion model is advantageous because it allows for rigorous uncertainty quantification. We expect that our excursion models are applicable more generally, e.g., for the modelling of higher-dimensional met-ocean time-series and spatial fields.

We model wind speed and significant wave height marginally conditional on directional covariates. However, we did not investigate the explicit effect of the directional components on the dependence models. Since, we remove the marginal effect of direction before modelling the dependence, we do not expect this covariate to have a significant impact on the dependence. However, it would be very interesting to adapt our models to be able to investigate this further in future research.

\appendix

\section{Reparameterization of EVAR}\label{paper4::app::reparam}
As opposed to inference for vector autoregressive models, we cannot estimate the EVAR parameters by least squares due to the presence of the $Y_{t,1}^{\bold{B}}$ term. Instead, we apply the inference methodology discussed in Section~\ref{paper4::sec::inference}. Not surprisingly, the parameter estimates $\hat{\Phi}^{(i)}$ for $i=1,\dots,k$ are highly intercorrelated because of the linear dependence between the components of $\bold{Y}_{t-1},\dots,\bold{Y}_{t-k}$. Reparameterization to reduce the correlation between parameter estimators is therefore attractive.

To reparameterize the model, we proceed as follows. First, we assume that the conditional extremes model is applicable to $Y_{t-i,j}$ conditional on $Y_{t-k,1}$ for each $i=0,\dots,k$ and $j=1,\dots,d$ apart from $(i,j)=(k,1)$, i.e., there exist parameters $\a_{i,j}\in[-1,1]$ and $\b_{i,j}<1$ such that 
\[\lim_{y\to\infty} \P\left(\frac{Y_{t-i,j} - \a_{i,j} y}{ y^{\b_{i,j}}} \leq x\ \Big|\ Y_{t-k,1}=y\right) = H_{i,j}(x), \]
where $H_{i,j}$ is a non-degenerate distribution function. Following the EVAR model~(\ref{paper4::eqn::evar_model_formulation}), we now must have
\[ \begin{aligned}
Y_{t+k,1}
&=  \Phi^{(1)}_{1,1} Y_{t+k-1,1} + \cdots + \Phi^{(1)}_{d,1}Y_{t+k-1,d} + \cdots + \Phi^{(k)}_{1,1} Y_{t,1} + \cdots + \Phi^{(k)}_{d,1} Y_{t,d} + Y_{t,1}^{B_1} \varepsilon_{t,1}\\
&= \left(\Phi^{(1)}_{1,1} \alpha_{k-1,1} + \cdots + \Phi^{(1)}_{d,1} \alpha_{k-1,d} + \cdots + \Phi^{(k)}_{1,1}  + \cdots + \Phi^{(k)}_{d,1}  \alpha_{0,d}\right) Y_{t,1} + o_p(Y_{t,1}) 
\end{aligned} \]
conditional on $Y_{t,1}>v$ as $v$ tends to infinity. On the other hand, we have $Y_{t+k,1}|(Y_{t,1}>v) = \alpha_{0,1} Y_{t,1} + o_p(Y_{t,1})$. So, 
\[
\alpha_{0,1} = \Phi^{(1)}_{1,1} \alpha_{k-1,1} + \cdots + \Phi^{(1)}_{d,1}\alpha_{k-1,d} + \cdots + \Phi^{(k)}_{1,1} \cdot 1 + \cdots + \Phi^{(k)}_{d,1} \alpha_{0,d},
\]
which explains the collinearity of the estimators. We now propose the following reparameterization $(\bold{B},\tilde{\Phi}^{(1)},\dots,\tilde{\Phi}^{(k)})$. For each $1\leq l\leq d$, we acquire $\tilde{\Phi}^{(k-i)}_{j,l}$, i.e., the $(j,l)$th element of $\tilde{\Phi}^{(k-i)}$, inductively with $0\leq i\leq k-1$, $1\leq j\leq d$. 
\[
\Phi_{j,l}^{(k-i)} = \begin{cases}  \hat{\a}_{0,l} + \tilde{\Phi}_{1,l}^{(k)},\ & \text{for}\ i=0,\ j = 1,\\ 
-\tilde{\Phi}^{(k-i)}_{j-1,l}\ \ \ \cdot  \hat{\a}_{i,j-1}/\hat{\a}_{i,j} + \tilde{\Phi}_{j,l}^{(k-i)},\  & \text{for}\ i=0,\dots,k-1,\ j=2,\dots,d,\   \text{conditional on}\ \tilde{\Phi}_{1,l}^{(k-i)}, \\
-\tilde{\Phi}^{(k-i+1)}_{d,l}\cdot{\hat{\a}_{i-1,d}}/{\hat{\a}_{i,1}} + \tilde{\Phi}_{1,l}^{(k-i)},\  &\text{for}\ i=1,\dots,k-1,\ j=1\ \text{conditional on}\ \tilde{\Phi}_{d,l}^{(k-i+1)}. \end{cases}\]
where $\hat{\a}_{i,j}$ is the maximum likelihood estimate for $\a_{i,j}$. Under this reparametrization, estimators of $\tilde{\Phi}_{j,k}^{(i)}$ are less correlated, which we demonstrated in unreported experiments comparing the dependence of the original and the reparameterized parameters using adaptive MCMC methodology \citep{roberts2009examples}. 

\bibliography{bibliography2}

\newcommand{\noop}[1]{}
\begin{thebibliography}{38}
\expandafter\ifx\csname natexlab\endcsname\relax\def\natexlab#1{#1}\fi
\providecommand{\url}[1]{\texttt{#1}}
\providecommand{\href}[2]{#2}
\providecommand{\path}[1]{#1}
\providecommand{\DOIprefix}{doi:}
\providecommand{\ArXivprefix}{arXiv:}
\providecommand{\URLprefix}{URL: }
\providecommand{\Pubmedprefix}{pmid:}
\providecommand{\doi}[1]{\href{http://dx.doi.org/#1}{\path{#1}}}
\providecommand{\Pubmed}[1]{\href{pmid:#1}{\path{#1}}}
\providecommand{\bibinfo}[2]{#2}
\ifx\xfnm\relax \def\xfnm[#1]{\unskip,\space#1}\fi
\bibitem[{Bauwens et~al.(2006)Bauwens, Laurent and Rombouts}]{bauwens2006}
\bibinfo{author}{Bauwens, L.}, \bibinfo{author}{Laurent, S.},
  \bibinfo{author}{Rombouts, J.V.}, \bibinfo{year}{2006}.
\newblock \bibinfo{title}{Multivariate {GARCH} models: a survey}.
\newblock \bibinfo{journal}{Journal of {A}pplied {E}conometrics}
  \bibinfo{volume}{21}, \bibinfo{pages}{79--109}.
\bibitem[{Bortot and Coles(2003)}]{bortot2003}
\bibinfo{author}{Bortot, P.}, \bibinfo{author}{Coles, S.},
  \bibinfo{year}{2003}.
\newblock \bibinfo{title}{Extremes of {M}arkov chains with tail switching
  potential}.
\newblock \bibinfo{journal}{Journal of the Royal Statistical Society: Series B
  (Statistical Methodology)} \bibinfo{volume}{65}, \bibinfo{pages}{851--867}.
\bibitem[{Bortot and Tawn(1998)}]{bortot1998}
\bibinfo{author}{Bortot, P.}, \bibinfo{author}{Tawn, J.A.},
  \bibinfo{year}{1998}.
\newblock \bibinfo{title}{Models for the extremes of {M}arkov chains}.
\newblock \bibinfo{journal}{Biometrika} \bibinfo{volume}{85},
  \bibinfo{pages}{851--867}.
\bibitem[{Chavez-Demoulin and Davison(2005)}]{chavez2005}
\bibinfo{author}{Chavez-Demoulin, V.}, \bibinfo{author}{Davison, A.C.},
  \bibinfo{year}{2005}.
\newblock \bibinfo{title}{Generalized additive modelling of sample extremes}.
\newblock \bibinfo{journal}{Journal of the Royal Statistical Society: Series C
  (Applied Statistics)} \bibinfo{volume}{54}, \bibinfo{pages}{207--222}.
\bibitem[{Coles and Tawn(1994)}]{coles1994statistical}
\bibinfo{author}{Coles, S.G.}, \bibinfo{author}{Tawn, J.A.},
  \bibinfo{year}{1994}.
\newblock \bibinfo{title}{Statistical methods for multivariate extremes: an
  application to structural design (with discussion)}.
\newblock \bibinfo{journal}{Journal of the Royal Statistical Society: Series C
  (Applied Statistics)} \bibinfo{volume}{43}, \bibinfo{pages}{1--31}.
\bibitem[{Davison and Smith(1990)}]{davison1990}
\bibinfo{author}{Davison, A.C.}, \bibinfo{author}{Smith, R.L.},
  \bibinfo{year}{1990}.
\newblock \bibinfo{title}{Models for exceedances over high thresholds (with
  discussion)}.
\newblock \bibinfo{journal}{Journal of the Royal Statistical Society: Series B
  (Methodological)} \bibinfo{volume}{52}, \bibinfo{pages}{393--425}.
\bibitem[{Eastoe and Tawn(2012)}]{eastoe2012}
\bibinfo{author}{Eastoe, E.F.}, \bibinfo{author}{Tawn, J.A.},
  \bibinfo{year}{2012}.
\newblock \bibinfo{title}{Modelling the distribution of the cluster maxima of
  exceedances of subasymptotic thresholds}.
\newblock \bibinfo{journal}{Biometrika} \bibinfo{volume}{99},
  \bibinfo{pages}{43--55}.
\bibitem[{Feld et~al.(2015)Feld, Randell, Wu, Ewans and Jonathan}]{feld2015}
\bibinfo{author}{Feld, G.}, \bibinfo{author}{Randell, D.}, \bibinfo{author}{Wu,
  Y.}, \bibinfo{author}{Ewans, K.}, \bibinfo{author}{Jonathan, P.},
  \bibinfo{year}{2015}.
\newblock \bibinfo{title}{Estimation of storm peak and intrastorm
  directional--seasonal design conditions in the {N}orth {S}ea}.
\newblock \bibinfo{journal}{Journal of Offshore Mechanics and Arctic
  Engineering} \bibinfo{volume}{137}.
\bibitem[{Gandy et~al.(2022)Gandy, Jana and Veraart}]{gandy2022}
\bibinfo{author}{Gandy, A.}, \bibinfo{author}{Jana, K.},
  \bibinfo{author}{Veraart, A.E.}, \bibinfo{year}{2022}.
\newblock \bibinfo{title}{Scoring predictions at extreme quantiles}.
\newblock \bibinfo{journal}{AStA Advances in Statistical Analysis}
  \bibinfo{volume}{106}, \bibinfo{pages}{527--544}.
\bibitem[{Heffernan and Tawn(2004)}]{heffernan2004conditional}
\bibinfo{author}{Heffernan, J.E.}, \bibinfo{author}{Tawn, J.A.},
  \bibinfo{year}{2004}.
\newblock \bibinfo{title}{A conditional approach for multivariate extreme
  values (with discussion)}.
\newblock \bibinfo{journal}{Journal of the Royal Statistical Society: Series B
  (Methodology)} \bibinfo{volume}{66}, \bibinfo{pages}{497--546}.
\bibitem[{Hilal et~al.(2011)Hilal, Poon and Tawn}]{hilal2011}
\bibinfo{author}{Hilal, S.}, \bibinfo{author}{Poon, S.H.},
  \bibinfo{author}{Tawn, J.A.}, \bibinfo{year}{2011}.
\newblock \bibinfo{title}{Hedging the black swan: Conditional
  heteroskedasticity and tail dependence in {S}\&{P}500 and {V}{I}{X}}.
\newblock \bibinfo{journal}{Journal of Banking \& Finance}
  \bibinfo{volume}{35}, \bibinfo{pages}{2374--2387}.
\bibitem[{Jan{\ss}en and Segers(2014)}]{janssen2014}
\bibinfo{author}{Jan{\ss}en, A.}, \bibinfo{author}{Segers, J.},
  \bibinfo{year}{2014}.
\newblock \bibinfo{title}{Markov tail chains}.
\newblock \bibinfo{journal}{Journal of Applied Probability}
  \bibinfo{volume}{51}, \bibinfo{pages}{1133--1153}.
\bibitem[{Joe(1997)}]{joe1997}
\bibinfo{author}{Joe, H.}, \bibinfo{year}{1997}.
\newblock \bibinfo{title}{Multivariate Models and Multivariate Dependence
  Concepts}.
\newblock \bibinfo{publisher}{CRC Press}.
\bibitem[{Keef et~al.(2013)Keef, Papastathopoulos and
  Tawn}]{keef2013estimation}
\bibinfo{author}{Keef, C.}, \bibinfo{author}{Papastathopoulos, I.},
  \bibinfo{author}{Tawn, J.A.}, \bibinfo{year}{2013}.
\newblock \bibinfo{title}{Estimation of the conditional distribution of a
  multivariate variable given that one of its components is large: Additional
  constraints for the {H}effernan and {T}awn model}.
\newblock \bibinfo{journal}{Journal of Multivariate Analysis}
  \bibinfo{volume}{115}, \bibinfo{pages}{396--404}.
\bibitem[{Leadbetter et~al.(1983)Leadbetter, Lindgren and
  Rootz{\'e}n}]{leadbetter1983}
\bibinfo{author}{Leadbetter, M.R.}, \bibinfo{author}{Lindgren, G.},
  \bibinfo{author}{Rootz{\'e}n, H.}, \bibinfo{year}{1983}.
\newblock \bibinfo{title}{Extremes and Related Properties of Random Sequences
  and Processes}.
\newblock \bibinfo{publisher}{New York: Springer-Verlag}.
\bibitem[{Ledford and Tawn(2003)}]{ledford2003diagnostics}
\bibinfo{author}{Ledford, A.W.}, \bibinfo{author}{Tawn, J.A.},
  \bibinfo{year}{2003}.
\newblock \bibinfo{title}{Diagnostics for dependence within time series
  extremes}.
\newblock \bibinfo{journal}{Journal of the Royal Statistical Society: Series B
  (Statistical Methodology)} \bibinfo{volume}{65}, \bibinfo{pages}{521--543}.
\bibitem[{Lugrin et~al.(2016)Lugrin, Davison and Tawn}]{lugrin2016bayesian}
\bibinfo{author}{Lugrin, T.}, \bibinfo{author}{Davison, A.C.},
  \bibinfo{author}{Tawn, J.A.}, \bibinfo{year}{2016}.
\newblock \bibinfo{title}{Bayesian uncertainty management in temporal
  dependence of extremes}.
\newblock \bibinfo{journal}{Extremes} \bibinfo{volume}{19},
  \bibinfo{pages}{491--515}.
\bibitem[{Ma and Swan(2020)}]{ma2020effective}
\bibinfo{author}{Ma, L.}, \bibinfo{author}{Swan, C.}, \bibinfo{year}{2020}.
\newblock \bibinfo{title}{The effective prediction of wave-in-deck loads}.
\newblock \bibinfo{journal}{Journal of Fluids and Structures}
  \bibinfo{volume}{95}, \bibinfo{pages}{102987}.
\bibitem[{Morison et~al.(1950)Morison, Johnson and Schaaf}]{morison1950force}
\bibinfo{author}{Morison, J.}, \bibinfo{author}{Johnson, J.},
  \bibinfo{author}{Schaaf, S.}, \bibinfo{year}{1950}.
\newblock \bibinfo{title}{The force exerted by surface waves on piles}.
\newblock \bibinfo{journal}{Journal of Petroleum Technology}
  \bibinfo{volume}{2}, \bibinfo{pages}{149--154}.
\bibitem[{Papastathopoulos et~al.(2023)Papastathopoulos, Casey and
  Tawn}]{papa2023}
\bibinfo{author}{Papastathopoulos, I.}, \bibinfo{author}{Casey, A.},
  \bibinfo{author}{Tawn, J.A.}, \bibinfo{year}{2023}.
\newblock \bibinfo{title}{Hidden tail chains and recurrence equations for
  dependence parameters associated with extremes of higher-order {M}arkov
  chains}.
\newblock \bibinfo{journal}{Submitted} .
\bibitem[{Papastathopoulos et~al.(2017)Papastathopoulos, Strokorb, Tawn and
  Butler}]{papa2017}
\bibinfo{author}{Papastathopoulos, I.}, \bibinfo{author}{Strokorb, K.},
  \bibinfo{author}{Tawn, J.A.}, \bibinfo{author}{Butler, A.},
  \bibinfo{year}{2017}.
\newblock \bibinfo{title}{Extreme events of {M}arkov chains}.
\newblock \bibinfo{journal}{Advances in Applied Probability}
  \bibinfo{volume}{49}, \bibinfo{pages}{134--161}.
\bibitem[{Perfekt(1997)}]{perfekt1997}
\bibinfo{author}{Perfekt, R.}, \bibinfo{year}{1997}.
\newblock \bibinfo{title}{Extreme value theory for a class of {M}arkov chains
  with values in $\mathbb{R}^d$}.
\newblock \bibinfo{journal}{Advances in Applied Probability}
  \bibinfo{volume}{29}, \bibinfo{pages}{138--164}.
\bibitem[{Randell et~al.(2015)Randell, Feld, Ewans and Jonathan}]{randell2015}
\bibinfo{author}{Randell, D.}, \bibinfo{author}{Feld, G.},
  \bibinfo{author}{Ewans, K.}, \bibinfo{author}{Jonathan, P.},
  \bibinfo{year}{2015}.
\newblock \bibinfo{title}{Distributions of return values for ocean wave
  characteristics in the {S}outh {C}hina {S}ea using directional--seasonal
  extreme value analysis}.
\newblock \bibinfo{journal}{Environmetrics} \bibinfo{volume}{26},
  \bibinfo{pages}{442--450}.
\bibitem[{Reistad et~al.(2009)Reistad, Breivik, Haakenstad, Aarnes and
  Furevik}]{WAM09}
\bibinfo{author}{Reistad, M.}, \bibinfo{author}{Breivik, O.},
  \bibinfo{author}{Haakenstad, H.}, \bibinfo{author}{Aarnes, O.J.},
  \bibinfo{author}{Furevik, B.R.}, \bibinfo{year}{2009}.
\newblock \bibinfo{title}{A high-resolution hindcast of wind and waves for the
  {N}orth {S}ea, the {N}orwegian {S}ea and the {B}arents {S}ea}.
\newblock \bibinfo{journal}{Norwegian Meteorological Institute} .
\bibitem[{Roberts and Rosenthal(2009)}]{roberts2009examples}
\bibinfo{author}{Roberts, G.O.}, \bibinfo{author}{Rosenthal, J.S.},
  \bibinfo{year}{2009}.
\newblock \bibinfo{title}{Examples of adaptive {MCMC}}.
\newblock \bibinfo{journal}{Journal of Computational and Graphical Statistics}
  \bibinfo{volume}{18}, \bibinfo{pages}{349--367}.
\bibitem[{Rootz{\'e}n(1988)}]{rootzen1988}
\bibinfo{author}{Rootz{\'e}n, H.}, \bibinfo{year}{1988}.
\newblock \bibinfo{title}{Maxima and exceedances of stationary {M}arkov
  chains}.
\newblock \bibinfo{journal}{Advances in {A}pplied {P}robability}
  \bibinfo{volume}{20}, \bibinfo{pages}{371--390}.
\bibitem[{Ross et~al.(2020)Ross, Astrup, Bitner-Gregersen, Bunn, Feld, Gouldby,
  Huseby, Liu, Randell, Vanem and Jonathan}]{ross2020}
\bibinfo{author}{Ross, E.}, \bibinfo{author}{Astrup, O.C.},
  \bibinfo{author}{Bitner-Gregersen, E.}, \bibinfo{author}{Bunn, N.},
  \bibinfo{author}{Feld, G.}, \bibinfo{author}{Gouldby, B.},
  \bibinfo{author}{Huseby, A.}, \bibinfo{author}{Liu, Y.},
  \bibinfo{author}{Randell, D.}, \bibinfo{author}{Vanem, E.},
  \bibinfo{author}{Jonathan, P.}, \bibinfo{year}{2020}.
\newblock \bibinfo{title}{On environmental contours for marine and coastal
  design}.
\newblock \bibinfo{journal}{Ocean Engineering} \bibinfo{volume}{195},
  \bibinfo{pages}{106194}.
\bibitem[{Shooter et~al.(2022)Shooter, Ross, Ribal, Young and
  Jonathan}]{shooter2021_2}
\bibinfo{author}{Shooter, R.}, \bibinfo{author}{Ross, E.},
  \bibinfo{author}{Ribal, A.}, \bibinfo{author}{Young, I.R.},
  \bibinfo{author}{Jonathan, P.}, \bibinfo{year}{2022}.
\newblock \bibinfo{title}{Multivariate spatial conditional extremes for extreme
  ocean environments}.
\newblock \bibinfo{journal}{Ocean Engineering} \bibinfo{volume}{247},
  \bibinfo{pages}{110647}.
\bibitem[{Simpson and Wadsworth(2021)}]{simpson2021}
\bibinfo{author}{Simpson, E.S.}, \bibinfo{author}{Wadsworth, J.L.},
  \bibinfo{year}{2021}.
\newblock \bibinfo{title}{Conditional modelling of spatio-temporal extremes for
  {R}ed {S}ea surface temperatures}.
\newblock \bibinfo{journal}{Spatial Statistics} \bibinfo{volume}{41},
  \bibinfo{pages}{100482}.
\bibitem[{Smith(1992)}]{smith1992}
\bibinfo{author}{Smith, R.L.}, \bibinfo{year}{1992}.
\newblock \bibinfo{title}{The extremal index for a {M}arkov chain}.
\newblock \bibinfo{journal}{Journal of {A}pplied {P}robability}
  \bibinfo{volume}{29}, \bibinfo{pages}{37--45}.
\bibitem[{Smith et~al.(1997)Smith, Tawn and Coles}]{smith1997}
\bibinfo{author}{Smith, R.L.}, \bibinfo{author}{Tawn, J.A.},
  \bibinfo{author}{Coles, S.G.}, \bibinfo{year}{1997}.
\newblock \bibinfo{title}{Markov chain models for threshold exceedances}.
\newblock \bibinfo{journal}{Biometrika} \bibinfo{volume}{84},
  \bibinfo{pages}{249--268}.
\bibitem[{Tendijck et~al.(2019)Tendijck, Ross, Randell and
  Jonathan}]{tendijck2019}
\bibinfo{author}{Tendijck, S.}, \bibinfo{author}{Ross, E.},
  \bibinfo{author}{Randell, D.}, \bibinfo{author}{Jonathan, P.},
  \bibinfo{year}{2019}.
\newblock \bibinfo{title}{A model for the directional evolution of severe ocean
  storms}.
\newblock \bibinfo{journal}{Environmetrics} \bibinfo{volume}{30},
  \bibinfo{pages}{e2541}.
\bibitem[{Tiao and Box(1981)}]{tiao1981}
\bibinfo{author}{Tiao, G.C.}, \bibinfo{author}{Box, G.E.},
  \bibinfo{year}{1981}.
\newblock \bibinfo{title}{Modeling multiple time series with applications}.
\newblock \bibinfo{journal}{Journal of the American Statistical Association}
  \bibinfo{volume}{76}, \bibinfo{pages}{802--816}.
\bibitem[{Tiao and Tsay(1989)}]{tiao1989}
\bibinfo{author}{Tiao, G.C.}, \bibinfo{author}{Tsay, R.S.},
  \bibinfo{year}{1989}.
\newblock \bibinfo{title}{Model specification in multivariate time series}.
\newblock \bibinfo{journal}{Journal of the Royal Statistical Society: Series B
  (Methodological)} \bibinfo{volume}{51}, \bibinfo{pages}{157--195}.
\bibitem[{Towe et~al.(2019)Towe, Tawn, Lamb and Sherlock}]{towe2019}
\bibinfo{author}{Towe, R.P.}, \bibinfo{author}{Tawn, J.A.},
  \bibinfo{author}{Lamb, R.}, \bibinfo{author}{Sherlock, C.G.},
  \bibinfo{year}{2019}.
\newblock \bibinfo{title}{Model-based inference of conditional extreme value
  distributions with hydrological applications}.
\newblock \bibinfo{journal}{Environmetrics} \bibinfo{volume}{30},
  \bibinfo{pages}{e2575}.
\bibitem[{Winter and Tawn(2016)}]{winter2016}
\bibinfo{author}{Winter, H.C.}, \bibinfo{author}{Tawn, J.A.},
  \bibinfo{year}{2016}.
\newblock \bibinfo{title}{Modelling heatwaves in central {F}rance: a case-study
  in extremal dependence}.
\newblock \bibinfo{journal}{Journal of the Royal Statistical Society: Series C
  (Applied Statistics)} \bibinfo{volume}{65}, \bibinfo{pages}{345--365}.
\bibitem[{Winter and Tawn(2017)}]{winter2017}
\bibinfo{author}{Winter, H.C.}, \bibinfo{author}{Tawn, J.A.},
  \bibinfo{year}{2017}.
\newblock \bibinfo{title}{$k$th-order {M}arkov extremal models for assessing
  heatwave risks}.
\newblock \bibinfo{journal}{Extremes} \bibinfo{volume}{20},
  \bibinfo{pages}{393--415}.
\bibitem[{Yun(2000)}]{yun2000}
\bibinfo{author}{Yun, S.}, \bibinfo{year}{2000}.
\newblock \bibinfo{title}{The distributions of cluster functionals of extreme
  events in a $d$th-order {M}arkov chain}.
\newblock \bibinfo{journal}{Journal of Applied Probability}
  \bibinfo{volume}{37}, \bibinfo{pages}{29--44}.

\end{thebibliography}
\bibliographystyle{elsarticle-harv}

\section*{Supplementary Information}

The supplementary information consists of a series of figures following the format of Figure~5 of the main text, for different model choices.

\begin{figure}[!htbp]
	\centering
	\includegraphics[width=\textwidth]{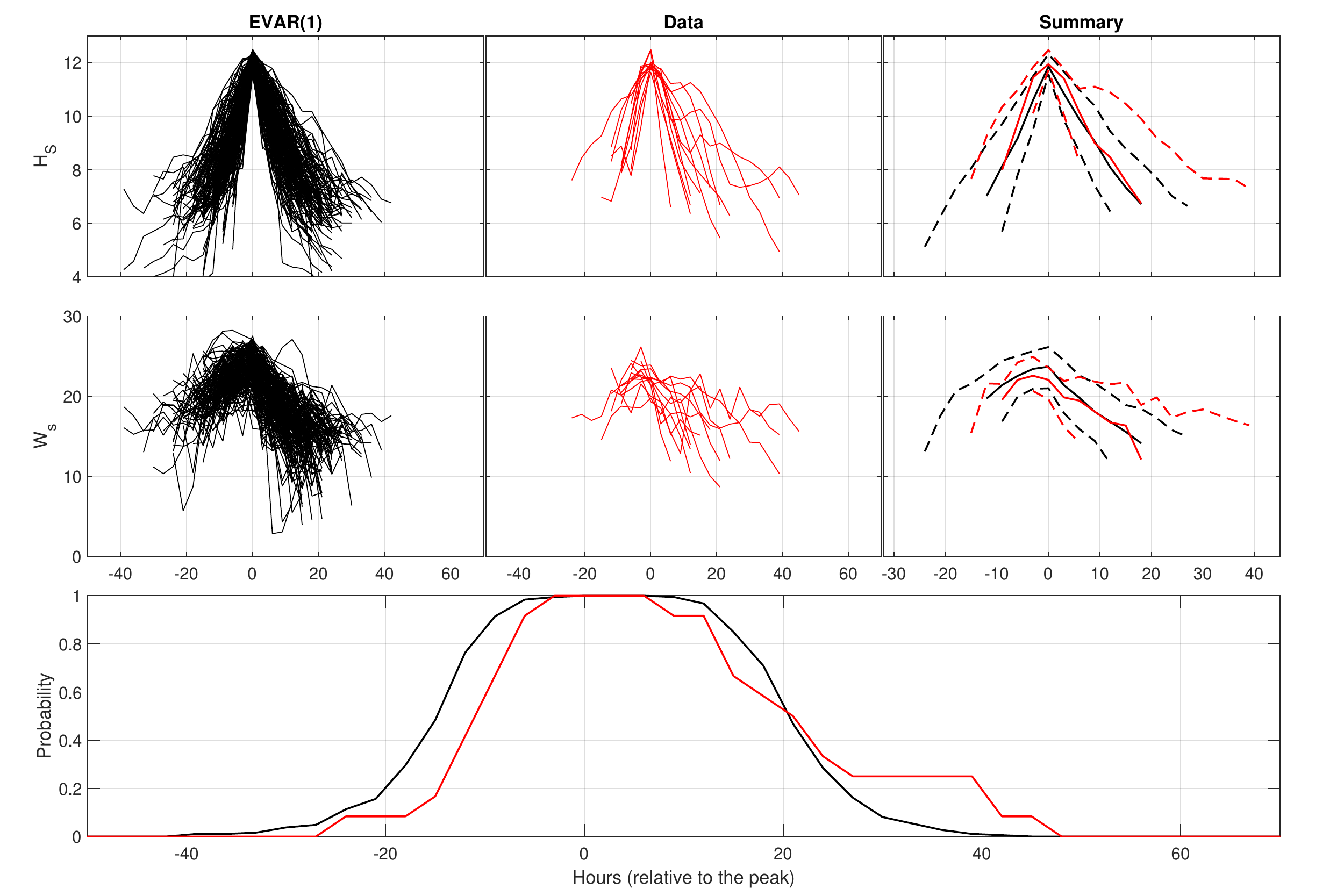}
	\caption{EVAR(1)}
\end{figure}
\begin{figure}[!htbp]
	\centering
	\includegraphics[width=\textwidth]{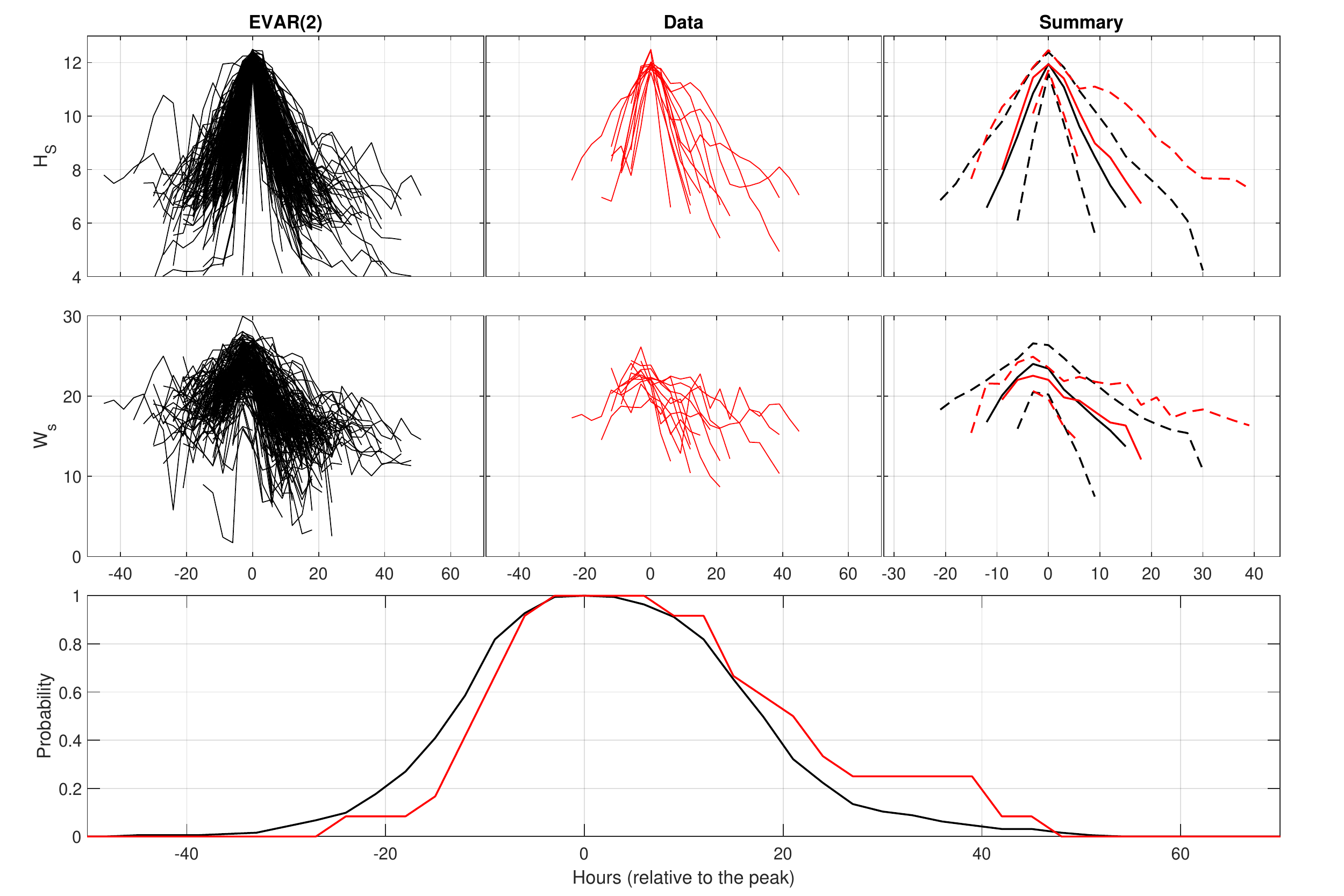}
	\caption{EVAR(2)}
\end{figure}
\begin{figure}[!htbp]
	\centering
	\includegraphics[width=\textwidth]{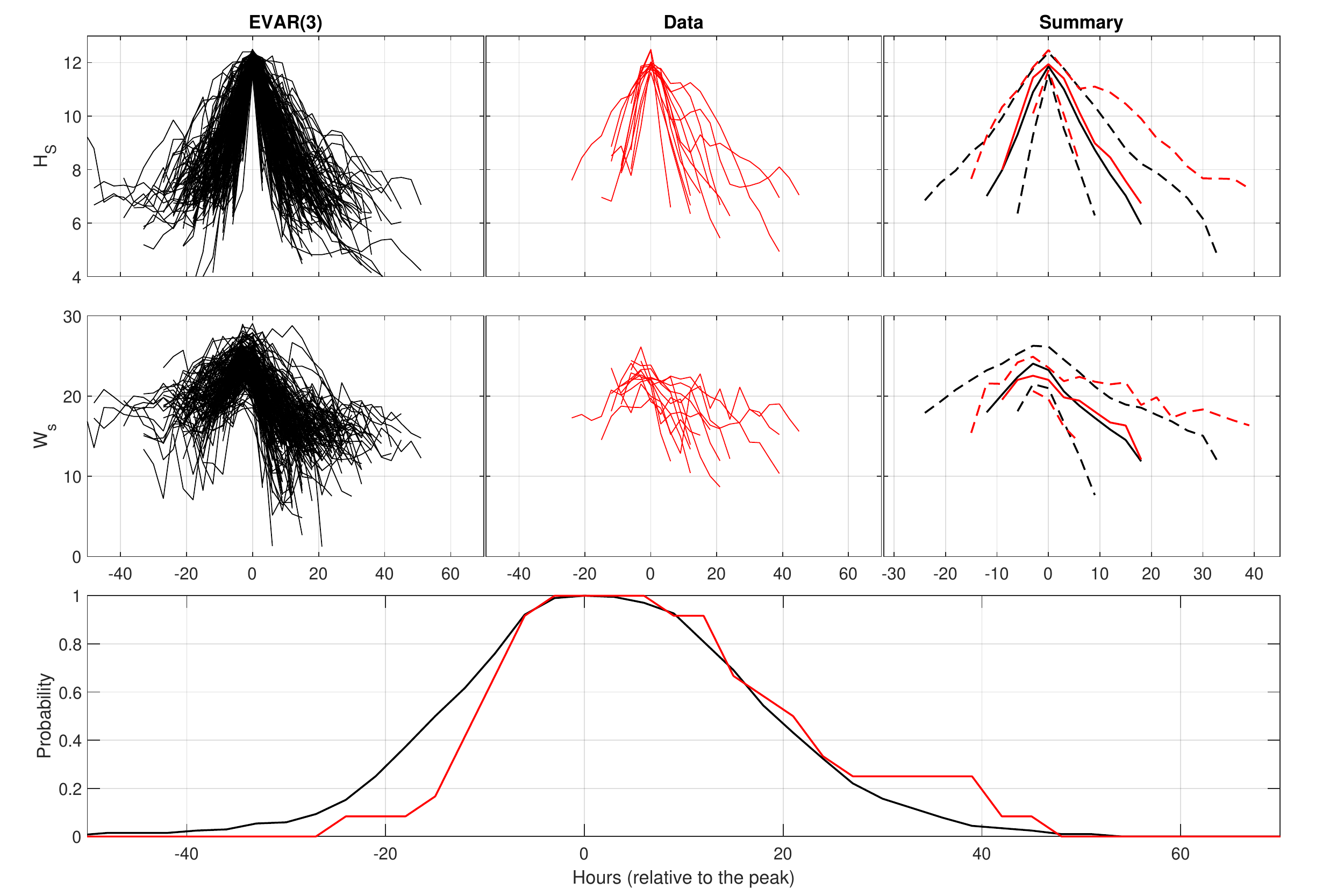}
	\caption{EVAR(3)}
\end{figure}
\begin{figure}[!htbp]
	\centering
	\includegraphics[width=\textwidth]{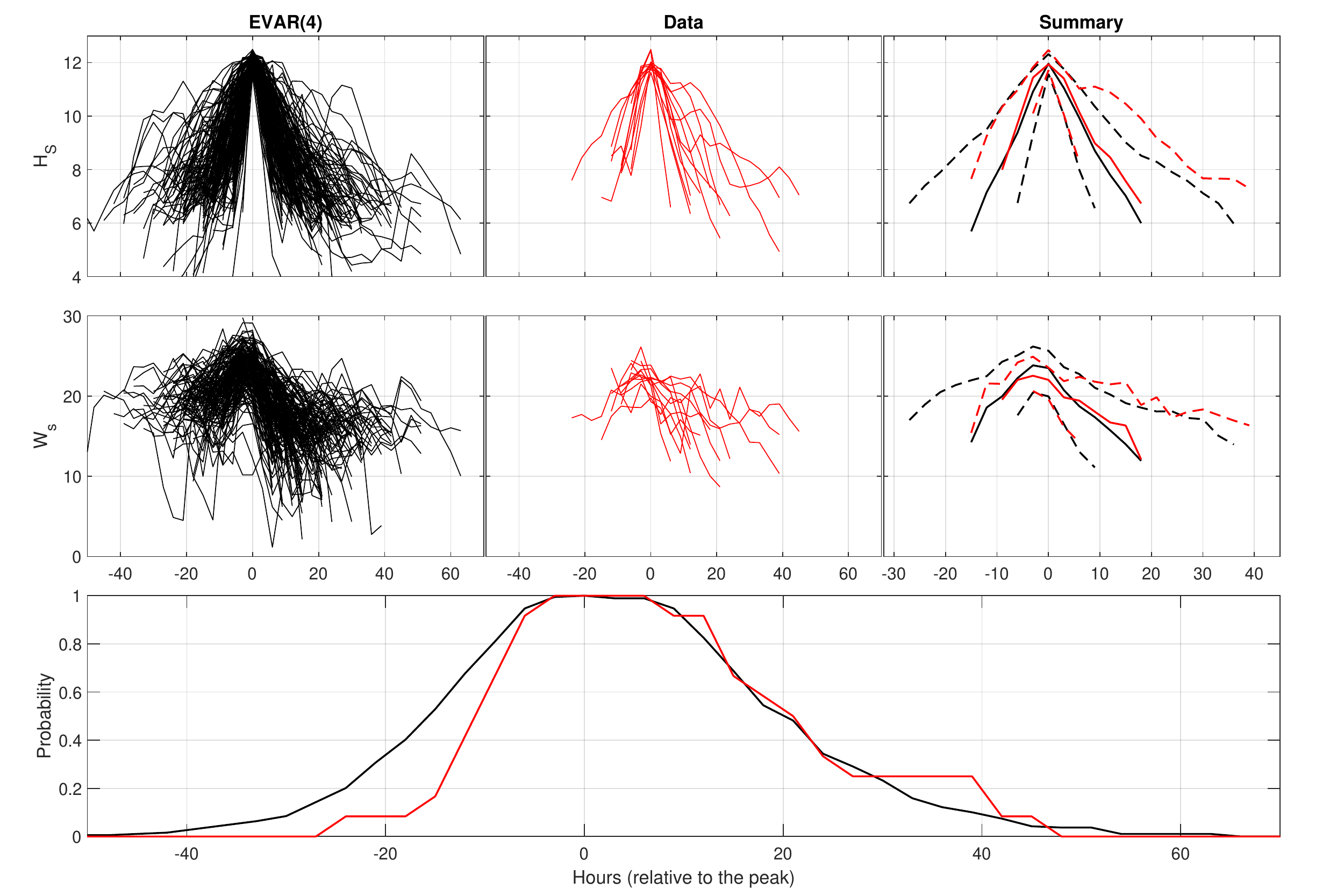}
	\caption{EVAR(4)}
\end{figure}
\begin{figure}[!htbp]
	\centering
	\includegraphics[width=\textwidth]{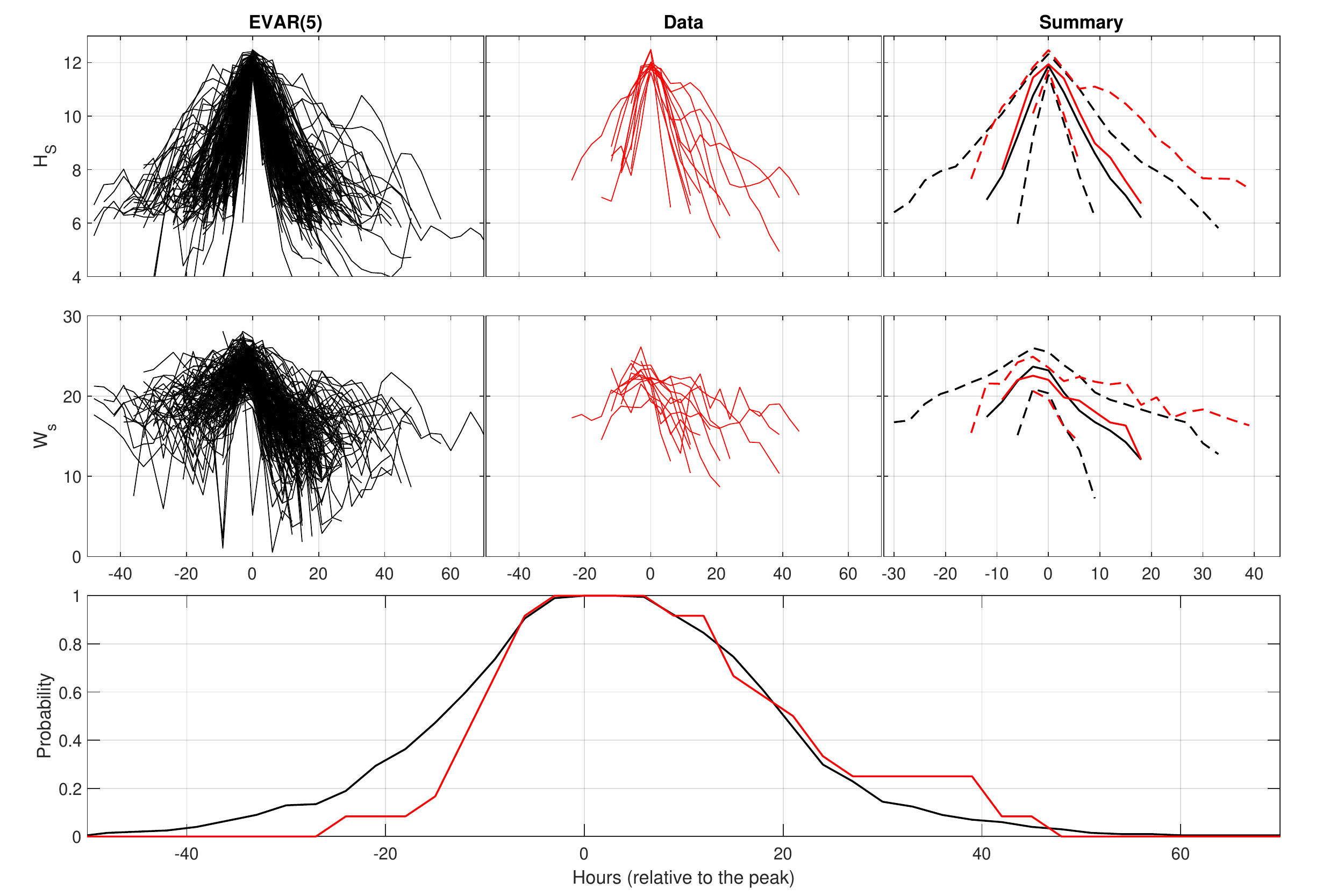}
	\caption{EVAR(5)}
\end{figure}
\begin{figure}[!htbp]
	\centering
	\includegraphics[width=\textwidth]{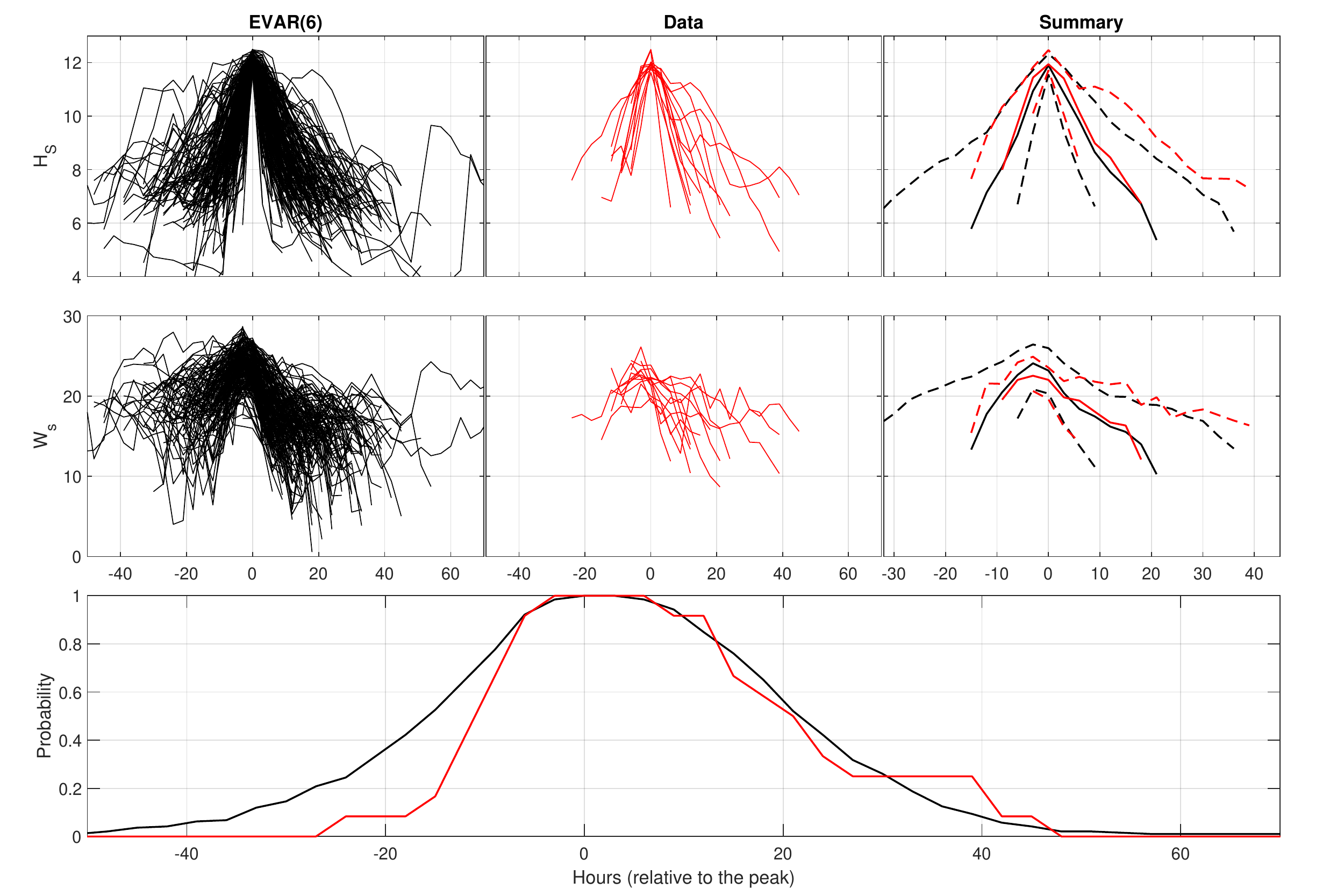}
	\caption{EVAR(6)}
\end{figure}

\begin{figure}[!htbp]
	\centering
	\includegraphics[width=\textwidth]{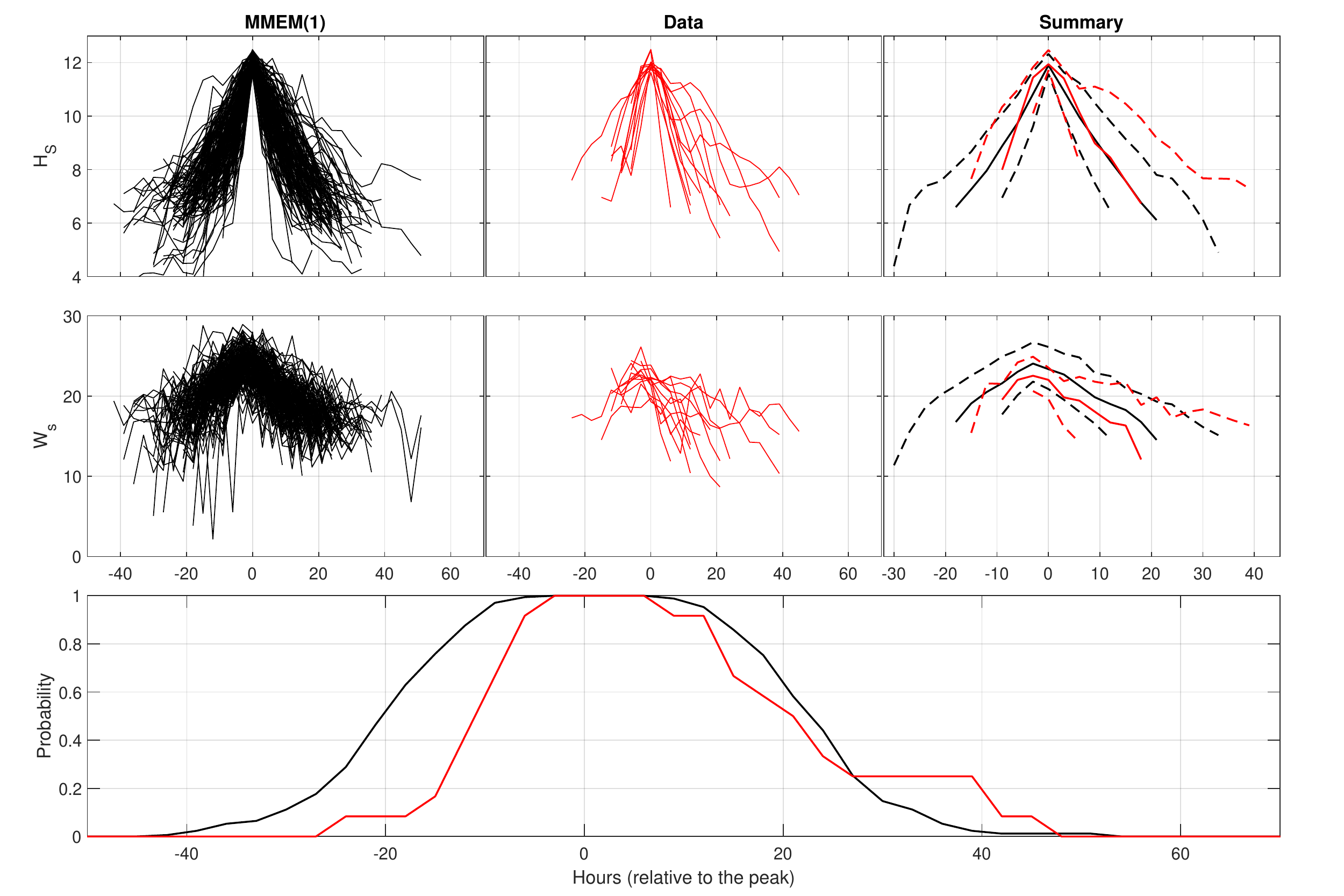}
	\caption{MMEM(1)}
\end{figure}
\begin{figure}[!htbp]
	\centering
	\includegraphics[width=\textwidth]{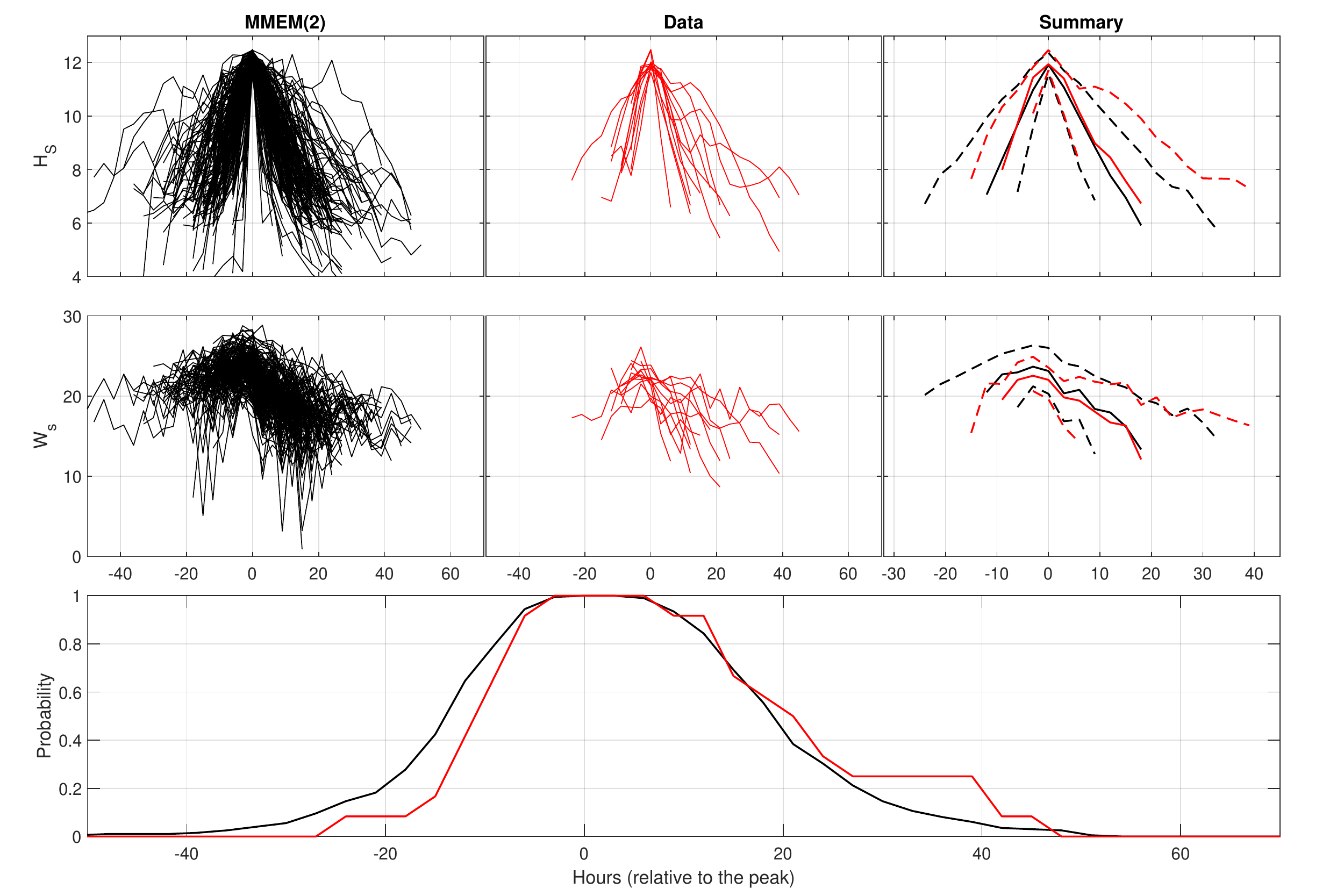}
	\caption{MMEM(2)}
\end{figure}
\begin{figure}[!htbp]
	\centering
	\includegraphics[width=\textwidth]{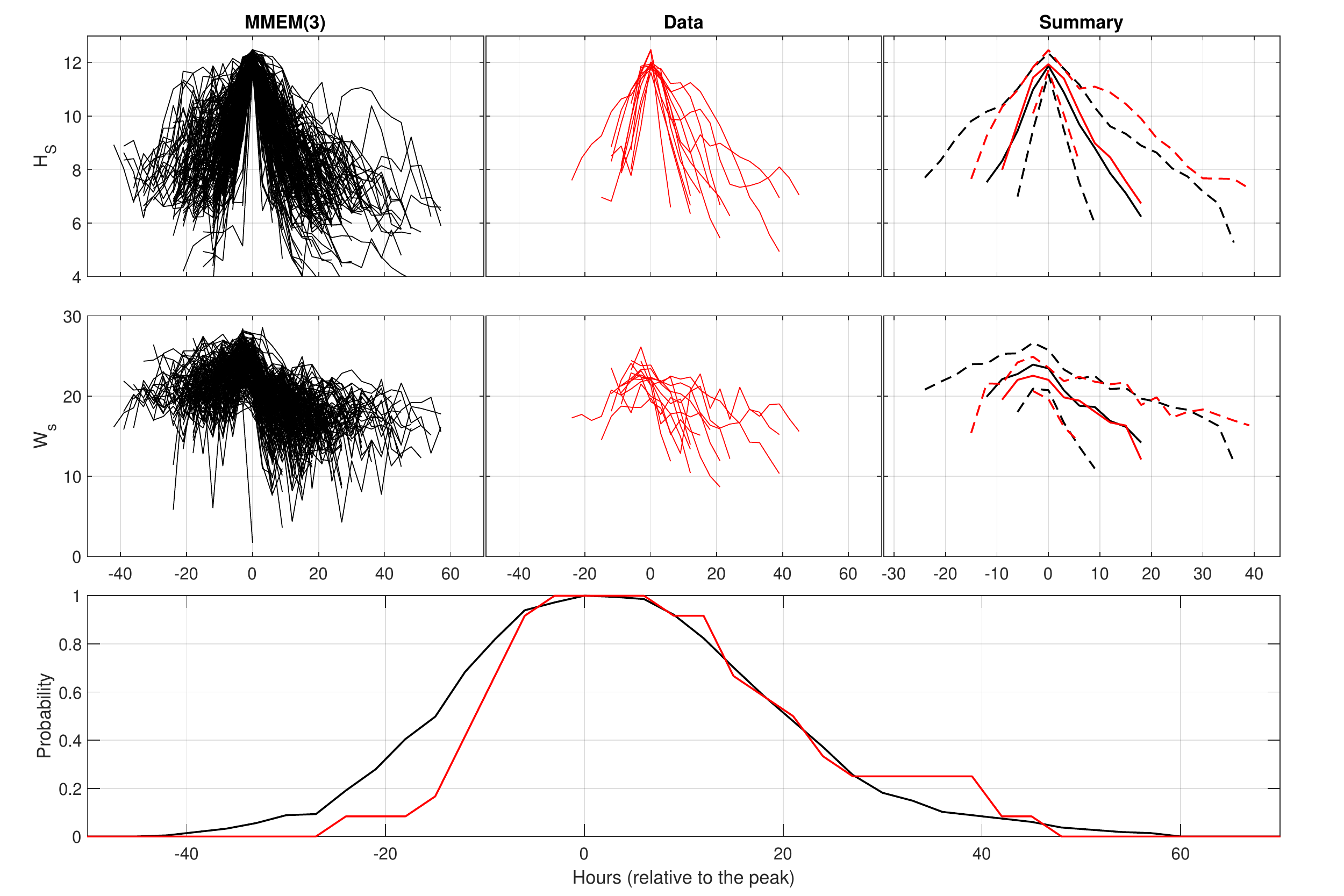}
	\caption{MMEM(3)}
\end{figure}
\begin{figure}[!htbp]
	\centering
	\includegraphics[width=\textwidth]{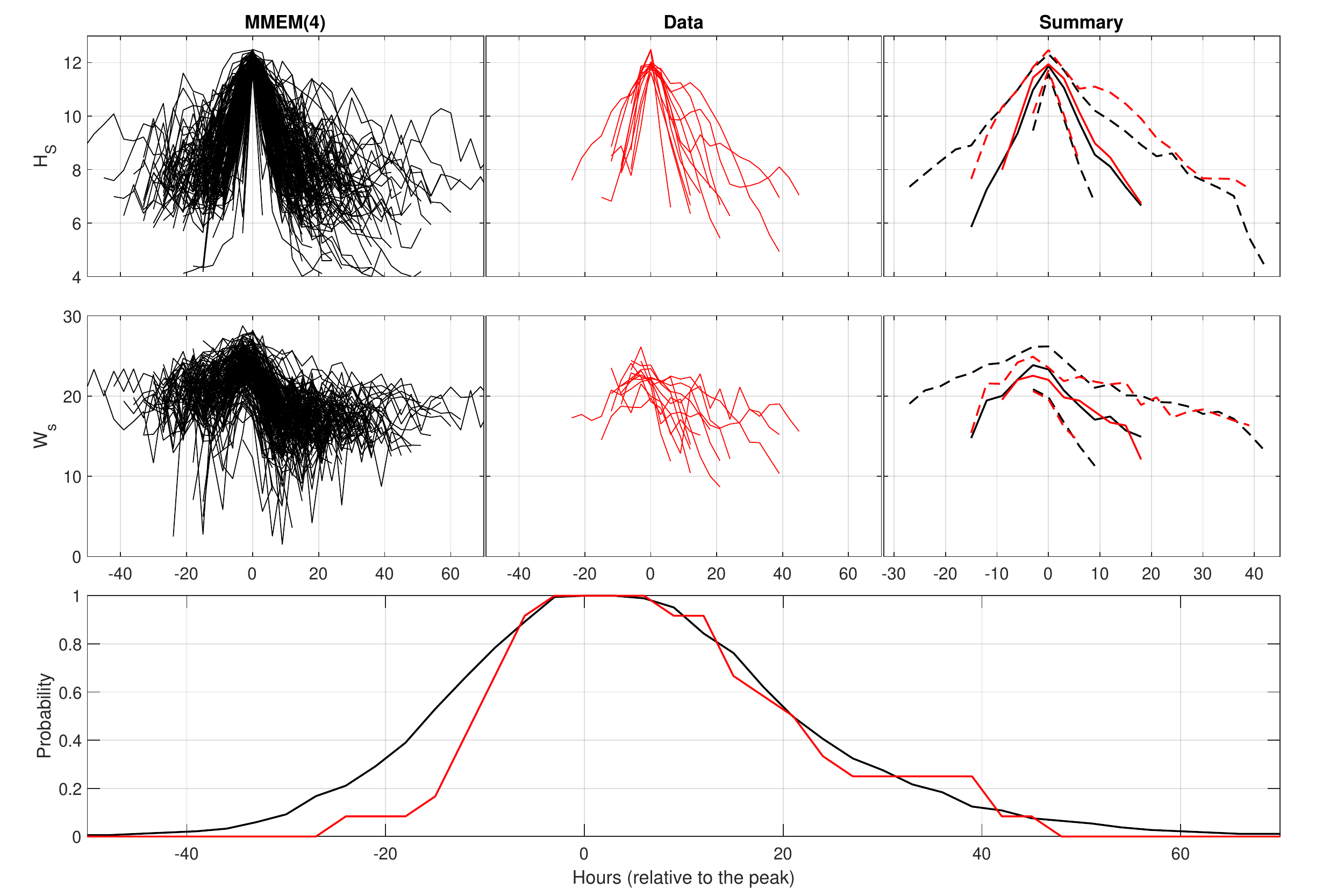}
	\caption{MMEM(4)}
\end{figure}
\begin{figure}[!htbp]
	\centering
	\includegraphics[width=\textwidth]{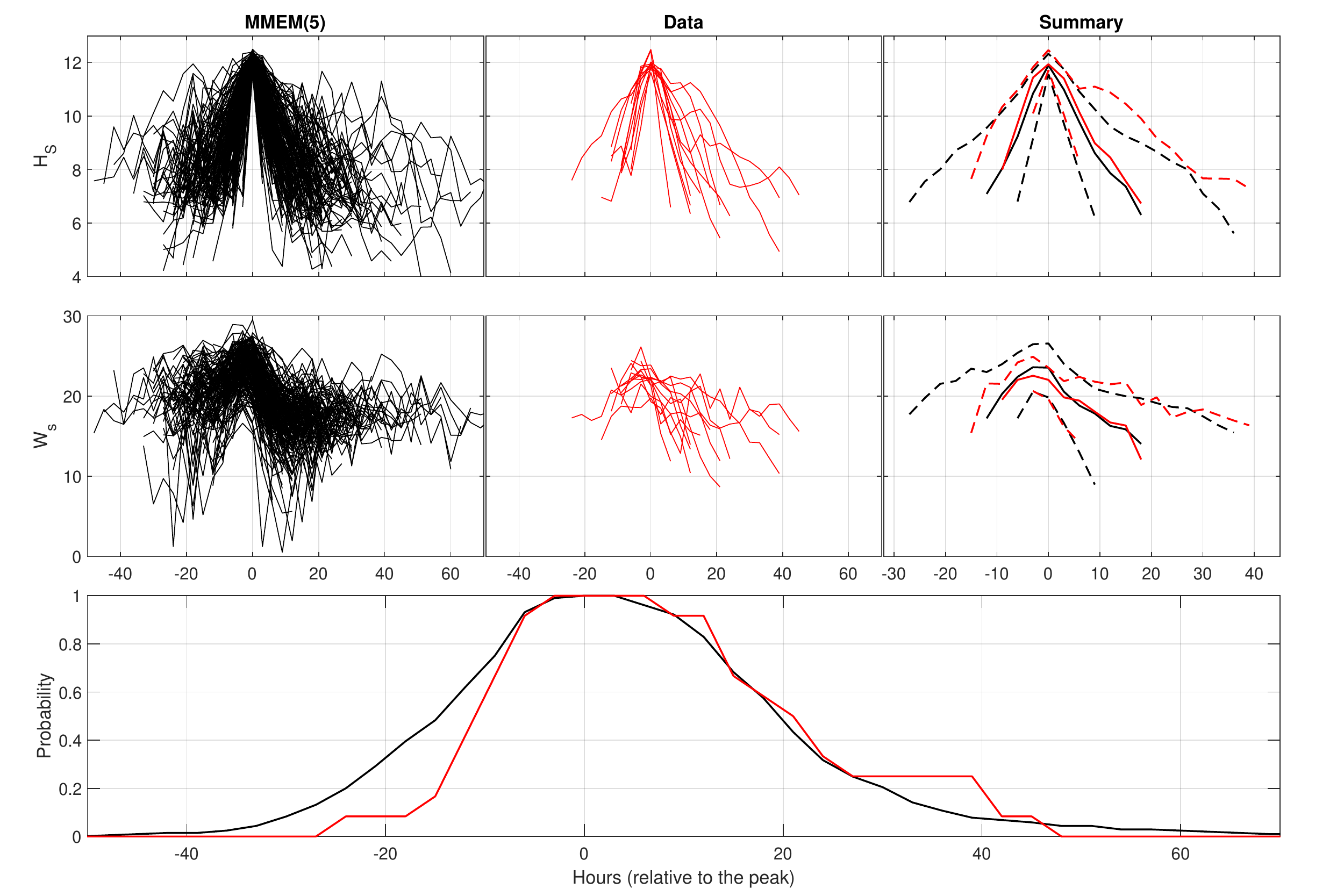}
	\caption{MMEM(5)}
\end{figure}
\begin{figure}[!htbp]
	\centering
	\includegraphics[width=\textwidth]{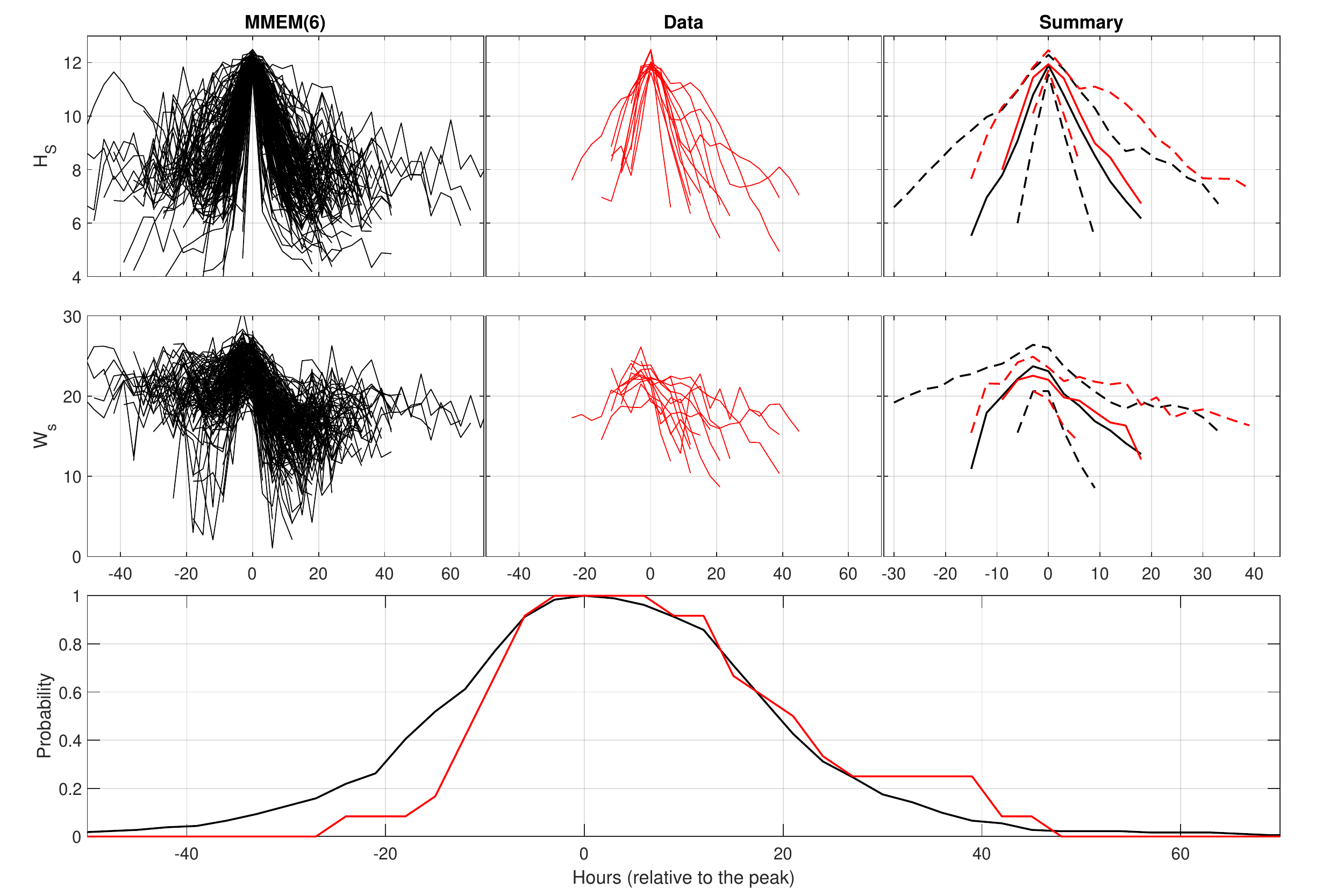}
	\caption{MMEM(6)}
\end{figure}
\begin{figure}[!htbp]
	\centering
	\includegraphics[width=\textwidth]{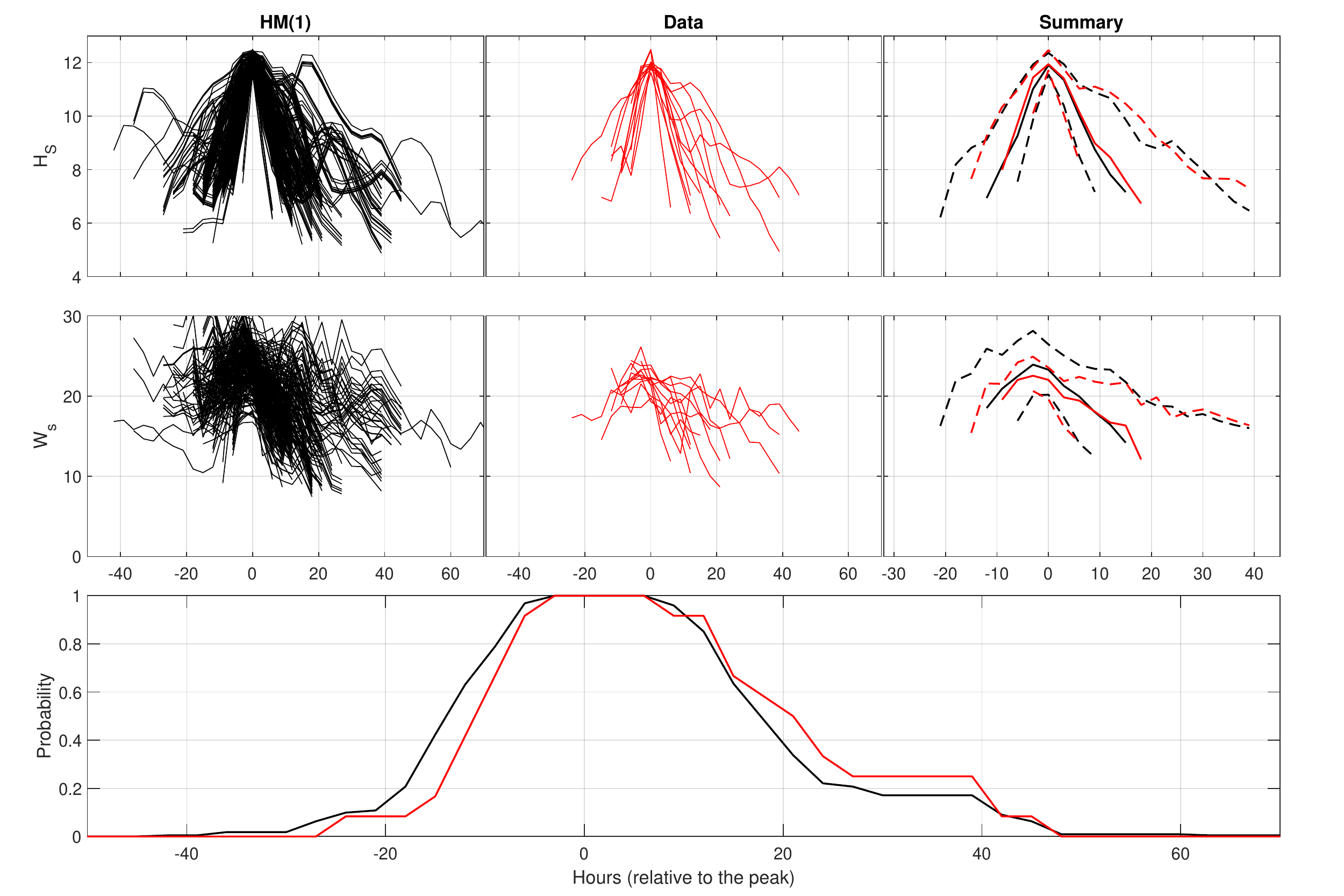}
	\caption{HM}
\end{figure}

\end{document}